\documentclass[sigconf,authorversion,nonacm,screen]{acmart}

\AtBeginDocument{%
  \providecommand\BibTeX{{%
    \normalfont B\kern-0.5em{\scshape i\kern-0.25em b}\kern-0.8em\TeX}}}

\usepackage{soul,color}

\usepackage{algorithm}
\usepackage{algpseudocode}
\usepackage{subcaption}
\usepackage{listings}
\usepackage[breakable,skins]{tcolorbox}
\usepackage{xspace}
\usepackage{pifont}
\usepackage{mathtools}
\usepackage{colortbl}
\usepackage{enumerate}

\definecolor{LightCyan}{RGB}{211,211,211}

\newcommand{\one}{\ding{182}\xspace}
\newcommand{\two}{\ding{183}\xspace}

\newcommand{\cmark}{\textcolor{blue}{\ding{51}}}%
\newcommand{\xmark}{\textcolor{red}{\ding{55}}}%

\newcommand{\sgx}{SGX\xspace}
\newcommand{\intelsgx}{Intel SGX\xspace}

\newcommand{\ecall}{\texttt{ECALL}\xspace}
\newcommand{\ocall}{\texttt{OCALL}\xspace}
\newcommand{\aex}{\texttt{AEX}\xspace}

\newcommand{\iozone}{Iozone\xspace}
\newcommand{\svm}{SVM\xspace}

\newcommand{\libcatena}{Blockchain\xspace}
\newcommand{\blockchain}{\libcatena\xspace}

\newcommand{\openssl}{OpenSSL\xspace}
\newcommand{\hashjoin}{HashJoin\xspace}
\newcommand{\btree}{B-Tree\xspace}
\newcommand{\bfs}{BFS\xspace}
\newcommand{\memcached}{Memcached\xspace}
\newcommand{\xsbench}{XSBench\xspace}
\newcommand{\lighttpd}{Lighttpd\xspace}
\newcommand{\pagerank}{PageRank\xspace}

\newcommand{\ncite}[1]{\citeauthor{#1} \cite{#1}}
\newcommand{\quotes}[1]{``#1''}

\lstdefinestyle{lstcode}{
language=C,
tabsize=4,
numbers=left,
showstringspaces=false,
basicstyle=\lstfont{black},
identifierstyle=\lstfont{black},
keywordstyle=\lstfont{blue},
numberstyle=\lstfont{black},
stringstyle=\lstfont{green},
commentstyle=\lstfont{brown!100},
emph={read,fsync,close,write,pread,pwrite,open,openat,lseek,CHUNK_DATA_SIZE_BYTES,HASH_SIZE,AES_KEY_SIZE
},
emphstyle={\lstfont{violet}},
breaklines=true,
frame=l,
captionpos=b,
    belowskip=1em,
aboveskip=1em
}

\usepackage{multirow}
\usepackage{array}
\usepackage{dblfloatfix}

\newcolumntype{C}[1]{>{\centering\arraybackslash}p{#1}}

\newcommand{\methodname}{SGXGauge\xspace}
\newcommand{\graphene}{GrapheneSGX\xspace}

\newcommand{\lmbenchsgx}{LMbench-SGX\xspace}
\newcommand{\nbenchsgx}{Nbench-SGX\xspace}

\newcommand{\lmbench}{LMbench\xspace}
\newcommand{\nbench}{Nbench\xspace}

\newcommand{\low}{Low\xspace}
\newcommand{\medium}{Medium\xspace}
\newcommand{\high}{High\xspace}

\newcommand{\vanilla}{\textit{Vanilla}\xspace}
\newcommand{\libos}{\textit{LibOS}\xspace}
\newcommand{\native}{\textit{Native}\xspace}

\tcbset{
        enhanced,
		top=0pt,left=0pt,right=5pt,bottom=0pt,
        boxrule=0.1pt,
		 breakable,
overlay broken = {
        \draw[line width=.1mm, red, rounded corners]
        (frame.north west) rectangle (frame.south east);},
       }

\title{A Comprehensive Benchmark Suite for \intelsgx}

\begin{document}

\frenchspacing

\author{Sandeep Kumar} 
 \affiliation{ 
\institution{School of Information Technology\\IIT Delhi}
\city{New Delhi}
\country{India}
}
\email{sandeep.kumar@cse.iitd.ac.in}

\author{Abhisek Panda} 
 \affiliation{ 
\institution{Department of Computer Science and Engineering, IIT Delhi}
\city{New Delhi}
\country{India}
}
\email{abhisek.panda@cse.iitd.ac.in}

\author{Smruti R. Sarangi} 
    \affiliation{ 
    \institution{Department of Computer Science and Engineering, IIT Delhi}
\city{New Delhi}
\country{India}
}
\email{srsarangi@cse.iitd.ac.in}

\begin{abstract}
Trusted execution environments (TEEs) such as \intelsgx facilitate the secure execution of an application on 
untrusted machines. Sadly, such environments suffer from serious limitations and performance overheads in terms of writing
back data to the main memory, their interaction
with the OS, and the ability to issue I/O instructions. There is thus a plethora of work that focuses on improving
the performance of such environments -- this necessitates the need for a standard, widely accepted benchmark suite (something
similar to SPEC and PARSEC). To the best of our knowledge, such a suite does not exist.

Our suite, \textit{\methodname}, contains a diverse
set of workloads such as blockchain codes, secure machine learning algorithms, lightweight web servers, secure key-value stores, etc. 
We thoroughly characterize the behavior of the benchmark suite on a native platform and on a platform
that uses a library OS-based shimming layer (\graphene). We observe that the most important metrics of interest are performance counters
related to paging, memory, and TLB accesses. There is an abrupt change in performance when the memory footprint starts to exceed
the size of the EPC size in \intelsgx, and the library OS does not add a significant overhead ($\approx \pm$ 10\%).
\end{abstract}

\maketitle

\section{Introduction}
\label{sec:intro}
Intel Secure Guard eXtension or \intelsgx~\cite{intelresearch, intelsgxexplained} has gained popularity in recent years as a way to securely execute an application on a remote, untrusted machine. The security of the application and data within \sgx, i.e.,  confidentiality, integrity, and freshness  are guaranteed by the hardware. 
The code and data within \intelsgx is even out of the reach of privileged system software such as the operating system and hypervisor. Recently, Microsoft Azure adopted \intelsgx to provide secure computing in their data centers~\cite{azure_sgx, sgx_datacenter}.

However, this protection comes at a cost. \intelsgx, to ensure security guarantees, puts certain restrictions on the applications running within it, such as no system calls, as the operating system is not a part of the trusted framework of \sgx~\cite{intelsgxexplained}. Therefore, few additional  and expensive  steps are required to enable system call support, which incurs additional performance overheads~\cite{portorshim,hotcalls}. 
Furthermore, \intelsgx reserves a portion of the main memory for its operations, which is managed by the hardware. However, this reserved memory is limited in size, and any application allocating more than the reserved memory, incurs a significant amount of performance overhead~\cite{intelsgxexplained,regaining_lost_seconds}.

Researchers have focused on alleviating this problem by proposing different mechanisms and workarounds to reduce the overheads~\cite{vault,scone,switchless_calls,eleos,hotcalls,regaining_lost_seconds}.
To show the benefits of their methods, researchers have resorted to manual porting of applications to \intelsgx~\cite{portorshim, sgxlapd_sgxnbench}. However, porting an application requires significant expertise and development effort~\cite{portorshim}. Also, the decision of which application to port is generally motivated by the ease of porting, and not necessarily by the gains accrued by doing so. 
Hence, there is no accepted, standard method for benchmarking \sgx-based systems primarily due to the ad hoc nature of 
workload creation.

The big picture is as follows. The workloads used to evaluate the efficiency of different methods for improving  \intelsgx vary across different proposals. Hence, it is not possible to compare the performance gains in one work with those of another in any meaningful manner.
Therefore, there is a need for a standard benchmark suite for \intelsgx, much like traditional benchmarks suites such as SPEC~\cite{spec2017} and PARSEC~\cite{parsec}.

A benchmark suite needs to thoroughly evaluate all the critical components of \intelsgx, and enable  performance comparison by setting a common denominator across different works.
Primarily, there are three sources of performance overheads in \intelsgx: encryption/decryption of the data in the reserved secure memory, the cost for accessing operating system services, and the additional time for swapping in data when an application has allocated more memory than the reserved memory~\cite{hotcalls, regaining_lost_seconds}. Prior works~\cite{ sgxlapd_sgxnbench,portorshim, sgxometer} in this field propose different benchmark suites~\cite{nbench_orig,lmbench} for evaluating \intelsgx. However, they only focus on the first two costs, ignoring the last one, which accounts for the maximum performance overhead~\cite{regaining_lost_seconds}.

We present \textit{\methodname} -- a comprehensive benchmark suite for \intelsgx. 
\methodname contains 10 real-world and synthetic benchmarks from different domains that thoroughly evaluate all the critical components of \intelsgx.
We use \methodname to evaluate \intelsgx in two different modes: \one \textit{native} mode where we port the benchmarks to \intelsgx, and \two \textit{shim} mode where we execute benchmarks in an environment where a  thin system software layer intercepts the system calls and intercedes with the OS on behalf of the application~\cite{graphenesgx,scone,panoply}. Such
shim layers are also known as library operating systems; they are gaining popularity because they significantly reduce the development time required to run an application on \sgx as compared to porting the same application to \sgx~\cite{portorshim}.
%
Our precise list of contributions are as follows.

\begin{enumerate}
\item We present \methodname, a benchmark suite for \intelsgx that thoroughly evaluates all of its components.
\item We stress test the impact of EPC on the performance of applications --- a crucial component that is missing from prior work.
\item We thoroughly evaluate the performance overhead incurred while executing an application with a library operating system.
\end{enumerate}

The rest of the paper is organized as follows: we discuss the required background for the paper in Section~\ref{sec:background}. In Section~\ref{sec:motivation_related_work} we discuss related work and the motivation for the paper. 
This is followed by a detailed overview of our benchmark suite, \methodname,  in Section~\ref{sec:benchmark_suite}. We discuss the evaluation of the benchmarks in Section~\ref{sec:evaluation}. 
We finally conclude in Section~\ref{sec:conclusion}.

\section{Background}
\label{sec:background}
In this section, we discuss the necessary background for the paper.

\subsection{\intelsgx}
Intel Secure Guard eXtension or SGX ensures the secure execution of an application either on a local or remote machine. It guarantees confidentiality, integrity, and freshness of the code and data running within it. Even privileged system software such as the operating system and hypervisor cannot affect its execution.

\sgx reserves a part of the system memory for its use at boot time. This reserved memory is known as the \textit{Processor Reserved Memory} or PRM~\cite{intelsgxexplained}.  Our system supports 128\,MB of PRM, and the rest of the discussion in the paper is based on this setting. The PRM is split into two regions that are used to store \one \sgx metadata and \two date/code of user applications, respectively. The latter is called the \textit{Enclave Page Cache} or EPC. The size of the EPC is 92\,MB, although \sgx supports applications that require more memory   (details in Section ~\ref{sec:epc}).
For every process, \sgx creates a trusted execution environment an {\em enclave}~\cite{intelsgxexplained}. 

The operating system cannot access the data within an enclave. However, an enclave still requires the operating system's support for setting it up, scheduling, context switching, page management, and cleanup. To enable this, the memory management of the enclave is done by the operating system.
Just before launching an enclave, the hardware checks the loaded binary for tampering by securely calculating its signature (hash) and matching it with the signature provided by the enclave's author.

\subsection{Enclave Page Cache}
\label{sec:epc}
The EPC is used to allocate memory for all the applications executing within \sgx. The  data in  the EPC is always in an encrypted form to prevent any snooping from   privileged system software such as the operating system. The data is decrypted when brought in to the LLC (last level cache) upon a CPU request. \intelsgx uses a dedicated hardware called the \textit{Memory Management Engine} or MEE for encrypting and decrypting the data.

Sadly, the size of the EPC is one of the major limitations of \sgx~\cite{regaining_lost_seconds}.
A typical modern application generally has a working set that is more than 92\,MB~\cite{working_set,working_set2}. In such cases, \sgx transparently evicts pages from the EPC to the untrusted memory, albeit in an encrypted form, to make space for the new data. When an application tries to access an evicted page, an EPC fault is raised, and \sgx brings the page back to the EPC~\cite{vault}.

An EPC fault is an expensive operation. \sgx encrypts and calculates the MAC (encrypted hash) of a page prior to an eviction. When the page is brought back, it needs to be decrypted and integrity checked before its use.
Our experiment found that evicting a page from the EPC takes on an average of 12,000 cycles.

\subsection{Enclave Transitions: {\ecall}s and {\ocall}s}
The security provided by \sgx comes at a cost. For security reasons, \sgx puts several restrictions on an executing enclave -- notably, it cannot make any system calls~\cite{intelsgxexplained}.

In the \intelsgx framework, the operating system is an untrusted entity, and hence, systems calls are restricted.
To make a system call, an enclave first exits the secure region by calling an \ocall (outside call) function.
After this, it makes the system call, collects its results, and returns to the secure region.
Similarly, an application from an unsecure region can call a function within an enclave by calling an \ecall (enclave call) function. 

\begin{figure}[!ht]
\centering
\includegraphics[width=.9\linewidth]{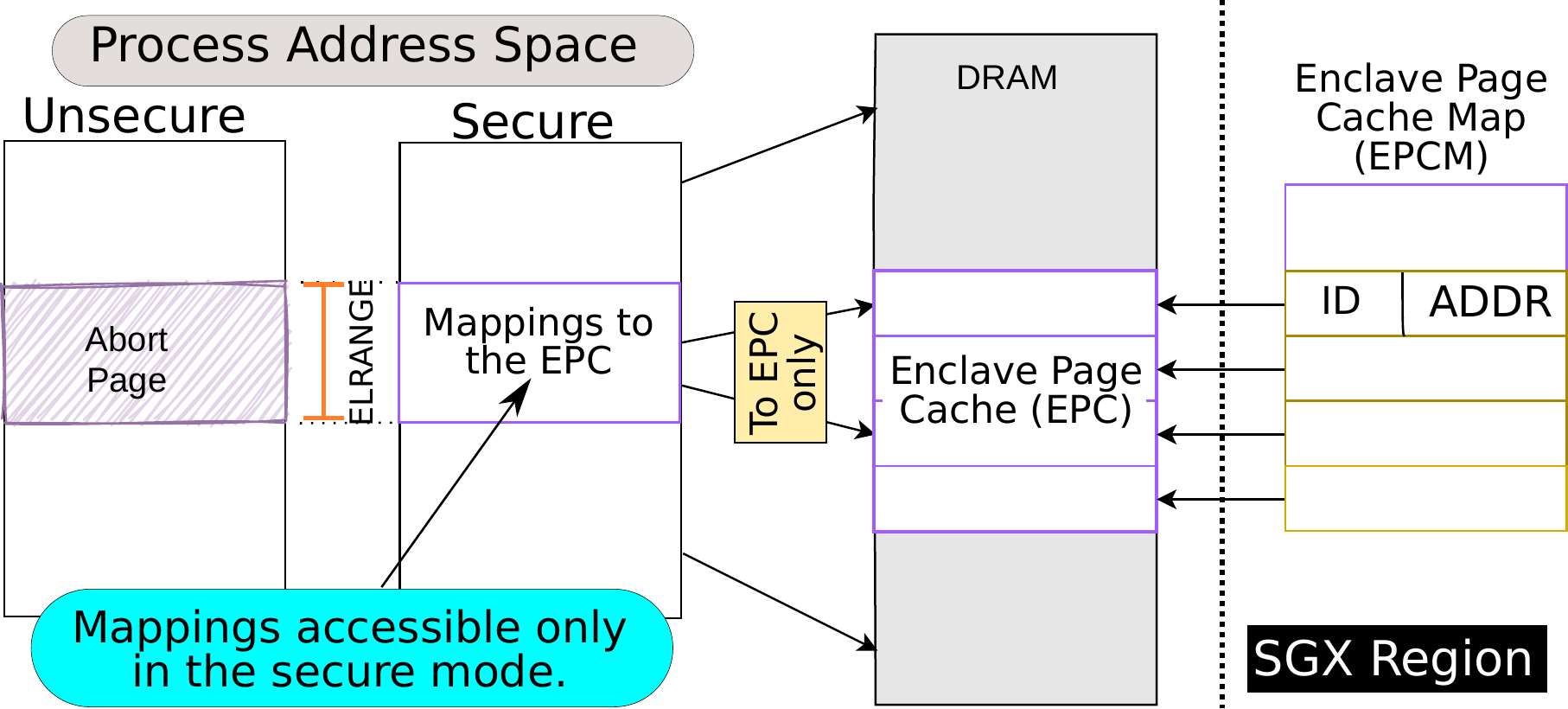}
\caption{Figure showing the relation between the address space, the EPC, and the EPCM.}
\label{fig:va_epc_elrange}
\end{figure}

During a transition from the secure region to the unsecure region, the TLB entries of the enclave are flushed due to security concerns~\cite{intelsgxexplained}. When the enclave returns, the TLB entries have to be populated again.
While adding a TLB entry to the TLB, if the entry points to an EPC page, it is first verified. For this purpose, \sgx maintains a table called the \emph{Enclave Page Cache Map} or EPCM~\cite{intelsgxexplained} in the secure region. The EPCM contains one entry for every page in the EPC. For each EPC page, the EPCM tracks its owner and the corresponding virtual address for which this page was allocated. These values are checked when a TLB entry for the corresponding page is being added to the TLB (see Figure ~\ref{fig:va_epc_elrange}).

Frequent enclave transitions affect the performance of an application due to context switches, TLB misses, and cache pollution. \ncite{hotcalls} show that the cost of calling an enclave function typically requires 17,000 cycles.

\subsection{Library Operating Systems}
Intel provides a software development kit (Intel \sgx SDK~\cite{intelsgxsdk}) to ease application development for \sgx. However, porting or writing an application for \sgx still requires an in-depth knowledge of the workings of \sgx and significant development effort~\cite{portorshim}. 

Due to this, many researchers have opted for executing an application using a shim layer built on top of \intelsgx -- also known as a {\em library operating system} or LibOS. A library operating system can execute an unmodified binary on \intelsgx; thus, saving on the high cost and effort of porting the application.
Scone~\cite{scone}, \graphene~\cite{graphenesgx}, and Panoply~\cite{panoply} are some examples of such systems. Researchers have reported that it can take up to 3-4 months for porting applications~\cite{portorshim}; this is also in line with our observations. Moreover, the task of verifying correct execution for all corner cases will take even more time. This is precisely why such shim layers need to be inevitably used and are fast becoming an inseparable part
of the \intelsgx stack. Even though they have their share of performance overheads, the sheer reduction in the development and verification effort makes them a necessary part of many deployments. No benchmark suite for \intelsgx can be oblivious of them.


\section{Related Work and Motivation}
\label{sec:motivation_related_work}
In this section, we discuss the related work in this area and the motivation for \methodname.

\subsection{Related Work}
Limited work has been done in this area, mainly due to the limitations of the \intelsgx framework and the engineering effort required to port an application to \sgx.

\subsubsection{\lmbenchsgx}
~\ncite{portorshim} in their work \textit{Port-or-Shim}, ported a part of \lmbench~\cite{lmbench} for \intelsgx and compared its performance against a shimmed version running within a library operating system. They also used \graphene for their evaluations. 
They point out that porting \lmbench to \sgx took months -- and that too after removing certain features from it~\cite{portorshim}. Whereas running a shimmed version of \lmbench on \graphene~\cite{graphenesgx} took a week of effort.
They specifically focus on the cost of the encryption/decryption and enclave transitions. They intentionally avoided EPC faults by ensuring that the amount of memory allocated to the benchmarks is less than the size of the EPC (92\,MB). 
They report that the performance of   \lmbenchsgx (ported version of \lmbench) and the version that runs within the library OS \graphene is the same -- this raises a question about whether porting was worth the effort.

\subsubsection{\nbenchsgx}
Apart from this, ~\ncite{sgxlapd_sgxnbench} proposed a method to prevent side-channel attacks in \intelsgx. They ported \nbench~\cite{nbench_orig} to \sgx to evaluate the effectiveness of their solution. However, the working set of the benchmarks was small and limited analyses were performed.

\leavevmode
\newline
Both \lmbenchsgx and \nbenchsgx (ported versions) are single-threaded benchmark suites~\cite{portorshim,sgxlapd_sgxnbench}. \lmbenchsgx mainly focuses on the memory bandwidth and the system call latencies. \nbenchsgx mostly contains CPU-intensive workloads and is designed to check the efficiency of integer and floating point operations of a CPU.
 Our suite is {\em far more comprehensive} in terms of its coverage (evaluated in Section~\ref{sec:evaluation}).

\ncite{sgxometer} also point out the issues with using \nbenchsgx for \intelsgx evaluation. They propose a CMake based framework to develop \sgx and non-\sgx applications from the same source. 
Apart from this, there are other proposals by ~\ncite{sgxperf}, ~\ncite{teemon}, ~\ncite{teeperf}, and ~\ncite{sgxtop} that propose methods to collect statistics about an executing secure application. This information can help a developer  debug a secure application or improve its performance. However, these are not benchmark suites; rather, they focus on the efficient profiling aspect for an executing enclave.

\subsection{Motivation}
Here, we discuss the motivation for \methodname. The system setup for experiments used here is listed in Table~\ref{tab:system_config}.

\subsubsection{\textbf{Experiment: Stressing the EPC}}

\begin{figure}
\centering
\includegraphics[width=.8\linewidth]{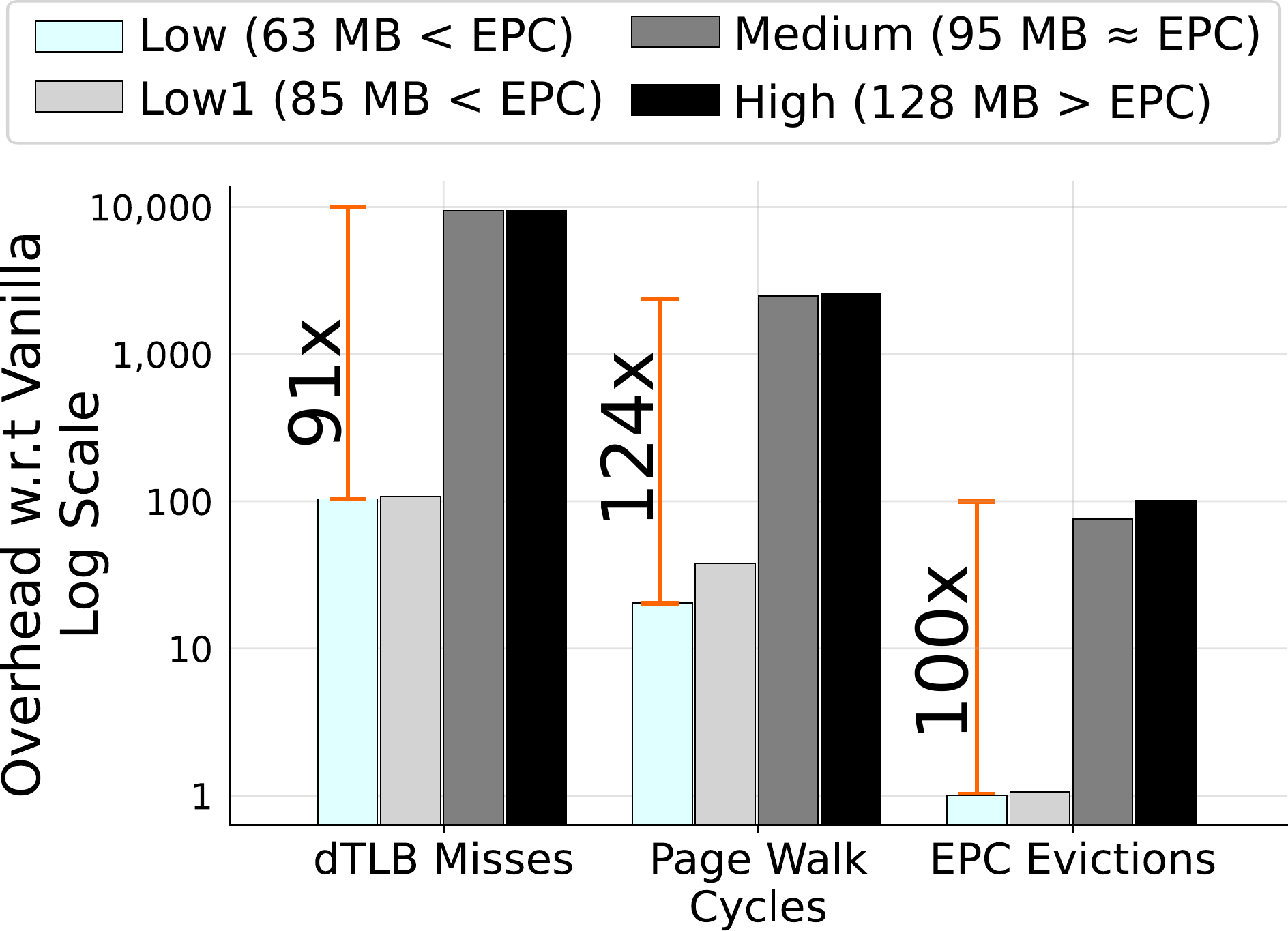}
\caption{Allocating beyond the EPC size increases the overhead. The baseline is a \vanilla (non-\sgx) setting with the same input size. For EPC evictions the baseline is the \low setting.}
\label{fig:motivation_epc}
\end{figure}

The limited amount of EPC memory is one of the biggest challenges in \sgx~\cite{regaining_lost_seconds, intelsgxexplained}. 
Due to the small size, EPC faults are a common event. Multiple instances of an enclave with a small memory footprint may also cause a number of EPC faults. This is because an enclave prior to its execution is loaded completely in the EPC to verify its content~\cite{everything_sgx_virtual,intelsgxexplained}.

As can be seen in Figure~\ref{fig:motivation_epc}, on crossing the EPC boundary the number of dTLB misses increases by 91$\times$, page walk cycles by more than 124$\times$, and EPC evictions by 100$\times$ as  compared to when the amount of memory is less than the EPC size. 
Hence, analyzing the impact of the  EPC size on the performance is crucial -- a fact completely ignored by \lmbenchsgx~\cite{portorshim} and \nbenchsgx~\cite{sgxlapd_sgxnbench}.

\subsubsection{\textbf{Experiment: Execution of multi-threaded benchmarks}} 

\begin{figure}
\centering
\includegraphics[width=.7\linewidth]{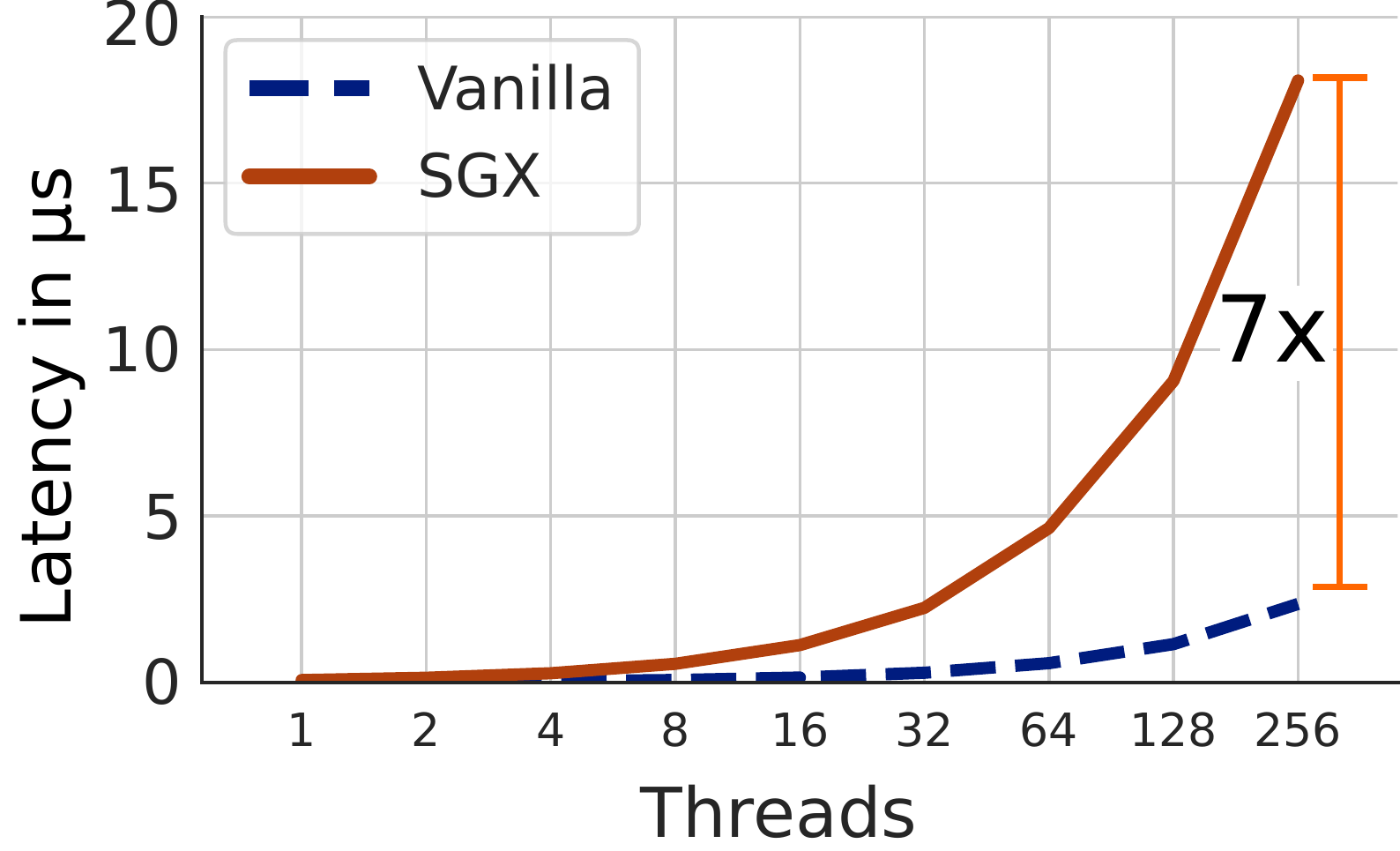}
\caption{The latency of the \lighttpd server increases with the number of concurrent accesses by up to 7$\times$ while running in \sgx and compared to a \vanilla (non-\sgx) execution.}
\label{fig:lighttpd}
\end{figure}

An application leverages the multiple cores on a modern system by using threads to speed up its operation. \intelsgx does not support thread creation inside the secure region; however, numerous threads can do an \ecall and execute the same function using the same global enclave ID~\cite{sgx_threads}. The overhead due to \intelsgx can change drastically based on the number of threads making the \ecall.
As shown in Figure~\ref{fig:lighttpd}, the latency of \lighttpd increases with the
number of threads (by 7$\times$).
Hence, it is crucial to capture the executions' characteristics in this setting also.
\nbenchsgx and \lmbenchsgx\,{\em do no contain} any multi-threaded benchmarks.

\subsubsection{\textbf{Experiment: Library Operating System}}

\begin{figure}
\centering
\includegraphics[width=.7\linewidth]{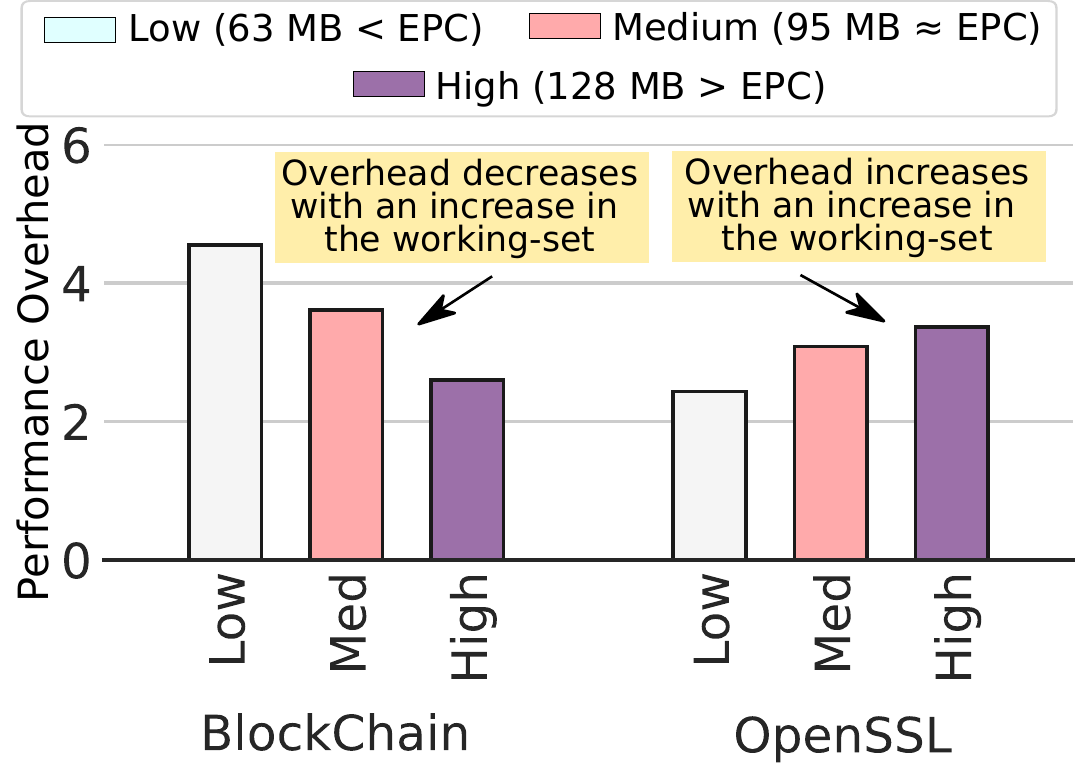}
\caption{A library operating system may affect the performance of an application in a positive or negative manner, depending on the characteristics of the application.}
\label{fig:libos}
\end{figure}

As noted in prior work~\cite{portorshim,sgxlapd_sgxnbench} and also by us, shimming an application is much easier than porting an application for \intelsgx -- in terms of development and verification effort. We believe that in the future a library operating system will be the primary way to execute an application on \intelsgx. Hence, it is essential to understand the behavioral changes between ported and shimmed applications. 
\textit{Port-or-Shim}~\cite{portorshim} also focuses on this problem, but with benchmarks that have a small working set (70\.MB). Our
observations are more comprehensive and also differ.
As shown in Figure~\ref{fig:libos}, the impact of a library operating system depends on the characteristics of the application
and thus needs to be rigorously studied.

\subsubsection{\textbf{Experiment: Real-world benchmarks}}

 Real world applications exhibit different phases during their execution. A typical pattern is that an application will read some data from the file system, process it, and then store the results. Micro-benchmarks such as \nbench~\cite{nbench_orig} lack this phase change behavior and thus do not represent a real-world scenario (details in Section~\ref{sec:evaluation}).\\
Let us summarize.

\begin{tcolorbox}[top=4pt,left=0pt,right=4pt,bottom=0pt]
\begin{enumerate}
\item Existing benchmark suites for \intelsgx~\cite{portorshim,sgxlapd_sgxnbench} are ported version of decade-old benchmarks that were designed to evaluate the CPU performance; they are not well suited to \sgx.
\item Using multiple threads changes the overheads of \intelsgx, hence it is necessary to also include multi-threaded benchmarks.
\item The performance impact of a library operating system depends on the characteristics of the application executing within it. Thus, there is a need for further study.
\item Modern applications show different phases during their execution. Therefore, it is necessary to  assess \intelsgx using real-world benchmarks.
\end{enumerate}
\end{tcolorbox}

\section{\methodname Benchmark Suite}
\label{sec:benchmark_suite}

The most important challenge in front of us was to find an appropriate set of workloads that 
need to be executed in a secure environment. 
This problem
has a degree of subjectivity. Nonetheless, we followed standard practice and restricted ourselves to workloads that have been used by highly cited works on \sgx in the recent past. We found the following workloads: blockchain related~\cite{blockchain_sgx,blockchain_sgx2,blockchain_sgx3,blockchain_sgx4,blockchain_sgx5}, protecting key-value pairs~\cite{keyvalue_sgx,keyvalue_sgx2,keyvalue_sgx3,keyvalue_sgx4}, securing databases~\cite{database_sgx,database_sgx2,database_sgx3,database_btree_sgx4}, protecting keys~\cite{password_sgx,password_sgx2}, securing a machine learning models~\cite{ml_sgx,ml_sgx2,ml_sgx3}, protecting network routing tables~\cite{hashtable_sgx}, securing communication~\cite{hash_sgx_signal}, graph traversals~\cite{graphtraversal_sgx,graphtraversal_sgx_2}, protecting web-servers~\cite{hotcalls,scone}, and HPC workloads~\cite{sgxl_gups_xsbench}.

The next task was to refine the set of workloads and choose an
appropriate set.
There are three main sources of overheads in \intelsgx: encryption/decryption, enclave transitions, and EPC faults (see Section~\ref{sec:background}).
While selecting workloads for \methodname, our primary aim was to ensure complete coverage of all the \intelsgx components. 
First, we selected some of the most commonly used workloads  such as \openssl~\cite{openssl,sgx_ssl} and \lighttpd~\cite{graphenesgx,hotcalls,occlum}. 
We then analyzed them and identified where they lack in terms of stressing the \sgx components. For e.g. both \openssl and \lighttpd do not stress the CPU much. Hence, to fulfil this criterion, we selected \libcatena workload which is a CPU-intensive and multi-threaded workload. However, while it stresses the CPU, it does not use a lot of memory. To ensure both the components are stressed, we opted for an HPC workload \xsbench, that was used by a prior work~\cite{sgxl_gups_xsbench} for similar purposes.

For selecting workloads that exclusively stress the EPC, we selected the following from prior work: \btree, \bfs, \hashjoin, and \pagerank. Each of them has different data access patterns. \btree is used commonly in databases for efficient lookups and has been used in \intelsgx setting also~\cite{database_btree_sgx4}. \bfs is used in protecting the control flow graph of an application~\cite{graphtraversal_sgx_2,flaas}. While \btree aims to minimize the number of nodes it visits, \bfs aims to visit all nodes efficiently. \hashjoin performs a number of hash table probing operations which is at the core of many systems~\cite{hashtable_sgx,hash_sgx_signal}.
\pagerank~\cite{pagerank_wiki} is a widely used workload for link analysis.
\svm is a machine learning (ML) workload that is CPU and memory-intensive. It runs multiple iterations over the same input data, a typical pattern of ML workloads.
We also discarded some workloads such as Redis~\cite{redis}, Fourier transform~\cite{rodinia}, License Managers~\cite{license3j}, GUPS~\cite{gups}, Nginx~\cite{nginx_paper}, etc. because they were similar to other workloads that were already chosen. 

\subsection{Evaluation Modes and Input Settings}
We execute the workloads in \methodname in different modes and different input settings to gain a better understanding of \intelsgx workings. Table~\ref{tab:conventions} list the different execution modes and input setting used in the paper.

\begin{table}[!ht]
	\centering
	\caption{Conventions used in the paper for discussion}
	\resizebox{\columnwidth}{!}{%
		\footnotesize
		\begin{tabular}{|l|p{6.5cm}|}
			\hline
			\rowcolor{LightCyan} \multicolumn{2}{|c|}{\textbf{Execution Modes}} \\ \hline 
			\vanilla              & An application executing without \intelsgx support.  \\ \hline
			\native              & An application executing within \intelsgx after it is ported to the \sgx framework. \\ \hline
			\libos & An application executing with \intelsgx in shimmed mode -- with the support of a LibOS (\graphene). \\ \hline
\hline
			\rowcolor{LightCyan} \multicolumn{2}{|c|}{\textbf{Input Settings}} \\ \hline 
\low & $Memory(footprint) < Size (EPC)$, wherever applicable \\
\hline
\medium & $Memory(footprint) \approx Size (EPC)$, wherever applicable\\
\hline
\high &   $Memory(footprint) > Size (EPC)$, wherever applicable\\
\hline
		\end{tabular}%
	}
	\label{tab:conventions}
\end{table}

\begin{table*}
	\centering
	\caption{Description of the workloads in \methodname along  with the specific settings used in the paper.}
{%
	\footnotesize
	\begin{tabular}{|l|l|C{1cm}|C{1cm}|C{1cm}|C{2cm}|C{1.8cm}|C{1.8cm}|C{1.8cm}|}
\hline
\textbf{S.No.} & \textbf{Workload}  & \textbf{\vanilla mode} &\textbf{\native mode} & \textbf{\libos mode} & \textbf{Property} & \textbf{Low ($<$ ~EPC)} & \textbf{Medium ($\approx$~EPC)} & \textbf{High ($>$ ~EPC)}
\\ \hline
1. &\libcatena \cite{libcatena}  & \cmark & \cmark & \cmark
& CPU/\ecall-intensive & Blocks 3 & Blocks 5 & Blocks 8
\\ \hline 
2. &OpenSSL \cite{openssl} & \cmark & \cmark & \cmark
& Data-intensive & File Size 76\,MB & File Size 88\,MB & File Size 151\,MB
\\ \hline 
3. &BTree~\cite{btree} & \cmark & \cmark & \cmark
& Data/CPU-intensive & Elements 1\,M& Elements 1.5\,M & Elements 2\,M
\\ \hline
4. &HashJoin~\cite{hashjoin} & \cmark & \cmark & \cmark
& Data/CPU-intensive & Data Table Size 61\,MB & Data Table Size 91\,MB & Data Table Size 122\,MB
\\ \hline \
5. &BFS~\cite{ligra} & \cmark & \cmark & \cmark
& Data-intensive & Nodes 70\,K \newline Edges 909\,K & Nodes 100\,K \newline Edges 1.3\,M & Nodes 150\,K\newline Edges 1.9\,M
\\ \hline 
6. &Pagerank~\cite{ligra}  & \cmark & \cmark & \cmark
& Data-intensive & Nodes 4500 \newline Edges 10.1\,M & Nodes 4750 \newline Edges 11.2\,M & Nodes 5000 \newline Edges 12.5\,M
\\ \hline 
7. &\memcached~\cite{memcached}& \cmark & \xmark & \cmark
& Data/\ecall-intensive & Records: 50K Operations:800K & Records: 100K Operations:800K & Records: 200K Operations:800K
\\ \hline 
8. &\xsbench~\cite{xsbench}& \cmark & \xmark & \cmark
& CPU-intensive & Points: 53K Lookups: 100& Points: 88K Lookups: 100 & Points: 768K Lookups: 100
\\ \hline 
9. &\lighttpd~\cite{lighttpd} & \cmark & \xmark & \cmark
& \ecall-intensive & Requests: 50K Threads: 16 & Requests: 60K Threads: 16 & Requests: 70K Threads: 16
\\ \hline 
10. &\svm~\cite{svm} & \cmark & \xmark & \cmark
& Data/CPU-intensive & Rows 4000 \newline Features 128 & Rows 6000 \newline Features 128 & Rows  10000 \newline Features 128
\\ \hline
	\end{tabular}
}

	\label{tab:workloads}
\end{table*}

\subsection{Workloads' Description}

\subsubsection{\textbf{Blockchain}~\cite{libcatena}:} A blockchain is a distributed ledger that does not require any central authority for its management. A blockchain is essentially a linked list of blocks. A block has a payload portion (contents of the block)
and the hash of contents of the previous block on the chain. Given that the chain is immutable, these hashes have a degree of
finality and permanence.
We vary the number of blocks in the chain to create different inputs for the workload (see Table ~\ref{tab:workloads}). 
The hash computation is the sensitive operation; hence, this operation is offloaded to \intelsgx. This function is called by many threads from the unsecure region resulting in many {~\ecall}s.

\subsubsection{\textbf{OpenSSL}~\cite{sgx_ssl,sgxometer}:} 
\openssl is a library that provides access to cryptographic primitives to developers. Our workload using Intel SGX-SSL~\cite{sgx_ssl} reads encrypted data from an input file and decrypts it within \sgx. Then, it performs a small compute-intensive task based on the content of the decrypted file. Finally, it encrypts the generated output and saves it in the untrusted filesystem.
This workload stresses the mechanisms that copy data from the unsecure memory region to the EPC and the EPC if the input file size is more than the EPC size.

\subsubsection{\textbf{\btree}~\cite{database_btree_sgx4}} The \btree data structure is used for an efficient organization of  data, specifically in database management systems. This enables efficient lookup in large databases and  applications of a similar nature -- a crucial feature in today's \quotes{big-data} world.
This workload creates a \btree consisting of a certain number of elements and performs multiple {\em find} operations on a randomly generated set of keys. This workload is also designed to stress the EPC and the paging system.

\subsubsection{{\textbf{\hashjoin}~\cite{hash_sgx_signal,hashtable_sgx}:}} The hash-join algorithm is used in modern databases to implement \quotes{equi-join}~\cite{equijoin_condition}. It has two phases: build and probe. Given two data tables, it first builds a hash table from the rows in the first table, and then probe it using the rows in the second table.
We vary the size of the first table and, in effect, vary the memory and compute-intensive nature of the workload.

\subsubsection{{\textbf{Breadth-First Search (BFS)}~\cite{graphtraversal_sgx,sabel_bfs}:}} The workload is a port from the implementation of the well-known breadth-first search algorithm used in the Rodinia benchmark suite~\cite{rodinia}. 
The input to the workload is an undirected graph. It first reads the input graph to the EPC and then traverses all the connected components in the graph.
This is primarily memory and compute-intensive workload. The computation overhead is a function of the number of nodes in the graph; the degree is at least 3. 

\subsubsection{{\textbf{Page Rank}~\cite{sgxl_gups_xsbench}:}} PageRank is used to rank web pages based on the popularity of pages that point to it.
The input to the workload is a connected directed graph represented in the adjacency list format with an out-degree of at least 1. The workload loads the graph into the EPC and builds an adjacency matrix of pages with a default initial rank for all. The workload then uses the number of out links of the page, previous rank, and the weight of the out neighbor pages to assign a new rank. This is repeated a fixed number of times.

\subsubsection{\textbf{\memcached}:~\cite{scone,memcached,hotcalls}} \memcached is an in-memory key-value store. It is used in production servers to cache hot data in memory. 
We use the popular YCSB~\cite{ycsb} workload to evaluate the performance of \memcached. YCSB first populates \memcached with a specified amount of data and then performs a specified set of (read or write) operations on those key-value pairs.

\subsubsection{\textbf{\xsbench}:~\cite{xsbench,sgxl_gups_xsbench}} \xsbench is a key computational kernel of the Monte Carlo neutron transport algorithm over a set of \quotes{nuclides} and \quotes{grid-points}~\cite{xsbench}. We vary the number of grid points to generate different input sizes for the workload  (see Table ~\ref{tab:workloads}).

\subsubsection{\textbf{\lighttpd}~\cite{lighttpd,hotcalls,graphenesgx}:} \lighttpd is a light-weight web server that is optimized for concurrent accesses. The server however runs on a single thread.
Our workload hosts a web-page of size 20\,KB (similar to \cite{hotcalls}). We use the \textit{ab} tool, which is a part of the Apache suite~\cite{ab_apache}, to make a certain number of requests to the \lighttpd server using concurrent threads (see Table~\ref{tab:workloads}). 

\subsubsection{\textbf{Support Vector Machine (SVM)}~\cite{libsvm}} SVM is a popular machine learning technique to classify the input data by projecting it into a higher dimensional space, and then using a linear combination of separating functions.
We implemented SVM using \textit{libSVM}~\cite{libsvm}, a library to implement SVM in C/C++ code. 

\subsection{Porting to \intelsgx}
\label{sec:porting_challenges}
\methodname contains 10 benchmarks. We have ported 6 of these to execute natively on \intelsgx ({\em native}). The other 4 are real-world benchmarks, which are
evaluated in \libos mode using \graphene (details in Section~\ref{sec:evaluation}). For these benchmarks, the engineering and verification
effort in creating a native \sgx port was prohibitive, and the benefits were not clear. Table~\ref{tab:conventions} summarizes this information and also
talks about the three execution settings (with different memory footprints): \low, \medium, and \high.

While porting an application to \intelsgx, the ideal case is to run the entire application within an enclave. However, this is not always possible due to the restrictions imposed by \intelsgx. In this case, a crucial function is typically moved to the enclave and is accessed via an \ecall. This is the standard practice~\cite{glamdring,enclavedom}.

We follow this approach while porting the applications. We completely ported \openssl, \bfs, \pagerank, \btree, and \hashjoin to \intelsgx. However, \libcatena uses multiple threads to speed up the hash finding process. \intelsgx does not support the creation of threads within an enclave~\cite{sgx_threads}. Nevertheless, multiple threads from the untrusted region can call the same \ecall function.
Hence, for \libcatena, we moved the hash function inside  \intelsgx; it is called from different threads from the main application which runs in the untrusted region.

\subsection{Running on \graphene}

To execute a binary on \graphene, we first need to define a \quotes{manifest} file. The manifest file contains the binary's location, list of libraries required, and the required input files. The parameters such as the enclave size and the threads to be used are also listed here. \graphene then processes this file and calculates the hash of all the required input files, which are then verified at the time of the execution.

\section{Evaluation}
\label{sec:evaluation}
Here, we discuss the performance of workloads in \methodname under different execution modes and with different input settings (see Table~\ref{tab:conventions}). We focus on the insights that are not listed in prior work~\cite{sgxometer,everything_sgx_virtual,portorshim,sgx_performance}, or where our observations differ from theirs.

\begin{table}[!ht]

\footnotesize
{%
\centering
\caption{System configuration}
\label{tab:system_config}
\begin{tabular}{lllll}
\hline 
\rowcolor{LightCyan} \multicolumn{5}{c}{\textbf{Hardware Settings}}                                                                                                              \\ \hline
\multicolumn{2}{|l|}{{Xeon E-2186G CPU, 3.80\ GHz}}                 &  \multicolumn{3}{l|}{Disk: 1\,TB (HDD)}     
\\ \hline
\multicolumn{5}{|l|}{CPUs: 1 Socket, 6 Cores, 2 HT} 
\\ \hline                    
\multicolumn{1}{|l|}{DRAM:  {32\,GB}}   &  \multicolumn{4}{l|}{ L1: 384\,KB, L2: 1536\,KB, L3: 12\,MB}       
\\  \hline
\rowcolor{LightCyan} \multicolumn{5}{c}{\textbf{System Settings}}                                                                                    
\textsl{}\\ \hline
\multicolumn{2}{|l|}{Linux kernel: 5.9}  & \multicolumn{2}{l|}{ASLR: Off}    & \multicolumn{1}{l|}{GCC: 9.3.0}           
     \\ \hline \multicolumn{2}{|l|}{DVFS: fixed frequency (performance)}      & \multicolumn{3}{l|}{Transparent Huge Pages: never}
\\ \hline
\rowcolor{LightCyan} \multicolumn{5}{c}{\textbf{SGX Settings}}                                                                                                        \\ \hline
\multicolumn{1}{|l|}{PRM: 128\,MB}  & \multicolumn{1}{l|}{Driver: 2.11}                  & \multicolumn{3}{l|}{SDK version: 2.13}       \\ \hline
\rowcolor{LightCyan} \multicolumn{5}{c}{\textbf{\graphene Settings}}                                                                                                        \\ \hline
\multicolumn{1}{|l|}{Enclave Size: 4\,GB}  & \multicolumn{1}{c|}{Threads: 16}                  & \multicolumn{3}{c|}{Internal Memory: 64\,MB}       \\ 
\bottomrule
\end{tabular}%
}

\end{table}

\subsection{Experimental Setup}
\label{sec:exp_setup}

The details of our evaluated system can be seen in Table~\ref{tab:system_config}.
We use the \graphene library operating system~\cite{graphenesgx} for our experiments in \libos mode.
Since \graphene is under constant development, we found that the performance of the code in the master branch is significantly better than its official release (v1.1). Hence, we used the code from the master branch of their GitHub repository\footnote{Commit ID: adf6269218dfa80aed276d57121a98e7b13b0f4e}.

\subsubsection{Instrumenting \intelsgx}
In order to instrument \sgx-related events, we added instrumentation code directly to the \intelsgx driver code. This approach has also been used in prior work~\cite{sgxtop,teemon}. We  identified crucial functions within the driver code that are called during different \sgx events, such as \texttt{sgx\_do\_fault()} (page fault handling function). Note that these functions do not execute within the secure world and thus can be easily instrumented.
We report the latencies of \sgx's crucial functions in Appendix ~\ref{appendix:latency}.

\begin{table}[]
\centering
\footnotesize
\label{tab:results_overview}

\caption{Overhead in system-related events. Avg. value of EPC evictions is reported when compared with the \vanilla mode. The overhead refers to the performance overhead (run time).}
\begin{tabular}{|p{.8cm}|p{.6cm}|C{.9cm}|C{.9cm}|C{.8cm}|C{.8cm}|C{1.1cm}|}
\hline
\rowcolor{LightCyan}  \multicolumn{7}{c}{{\native Mode w.r.t \vanilla (6 workloads)}}                            \\ \hline
             & \textbf{Over- head}   & \textbf{dTLB misses} &\textbf{ Walk Cycles} &\textbf{ Stall cycles} & \textbf{LLC misses} & \textbf{EPC Evictions} \\ \hline
\textbf{Low}         & 2.0$\times$   & 8.38$\times$      & 29.7$\times$               & 2.5$\times$         & 1.8$\times$&21.5\,K           \\ \hline
\textbf{Medium}      & 3.0$\times$ & 14.6$\times$      & 57.0$\times$              & 5.3$\times$         & 2.0$\times$&49.6\,K          \\ \hline
\textbf{High}        & 3.4$\times$ & 17.48$\times$      & 59.1$\times$              & 6.4$\times$         & 3.0$\times$&79.6\,K          \\ \hline
\rowcolor{LightCyan}  \multicolumn{7}{c}{{\libos Mode w.r.t \vanilla (10 workloads)}}                            \\ \hline
\textbf{Low}         & 2.03$\times$   & 40.6$\times$      & 517$\times$               & 114$\times$         &24$\times$  &    796\,K \\ \hline
\textbf{Medium}      & 3.13$\times$ & 59.7$\times$      & 724$\times$              & 146$\times$         & 18.5$\times$  &    1,792\,K  \\ \hline
\textbf{High}        & 3.7$\times$ & 44.0$\times$      & 113$\times$              & 12.7$\times$         & 15.5$\times$   &   2,255\,K    \\ \hline
\rowcolor{LightCyan}  \multicolumn{7}{c}{{\libos Mode w.r.t Native (6 workloads)}}                            \\ \hline
\textbf{Low}         & 1.03$\times$   & 3.3$\times$      & 5.1$\times$               & 8.3$\times$         & 9.3$\times$  &    75$\times$     \\ \hline
\textbf{Medium}      & 1.03$\times$ & 2.7$\times$      & 4.0$\times$              & 7.9$\times$         & 9.2$\times$  &    68$\times$  \\ \hline
\textbf{High}        & 0.9$\times$ & 2.0$\times$      & 3.0$\times$              & 5.9$\times$         & 7.2$\times$   &   45$\times$    \\ \hline
\end{tabular}%

\end{table}


\subsection{Evaluation Plan}
We take the following approach for evaluation.
\begin{itemize}
\item \textbf{\native mode performance:} We analyze the impact of \intelsgx on the applications executing natively on it for different input sizes.
\item \textbf{\libos mode performance}: We study the overheads that are introduced while executing on \intelsgx using a library operating system.
\item \textbf{\native mode vs. \libos mode}: We compare the performance of the \native and \libos execution modes.
\end{itemize}

Table ~\ref{tab:workloads} shows an overview of the evaluation results. The geometric mean value is computed across at least 10 executions (seen to be enough).

\begin{figure}
\centering
\begin{subfigure}[t]{\linewidth}
\includegraphics[width=\linewidth]{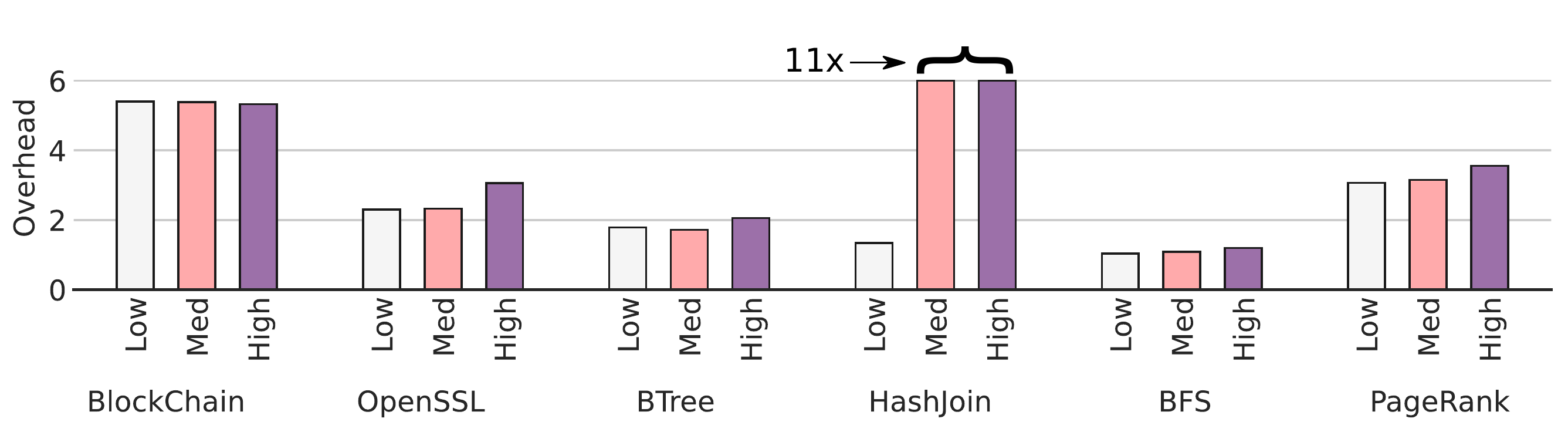}
\caption{Runtime overhead in \native mode.}
\label{fig:native_overhead_sftime_mod}
\end{subfigure}
\hfil
\begin{subfigure}[t]{\linewidth}
\includegraphics[width=\linewidth]{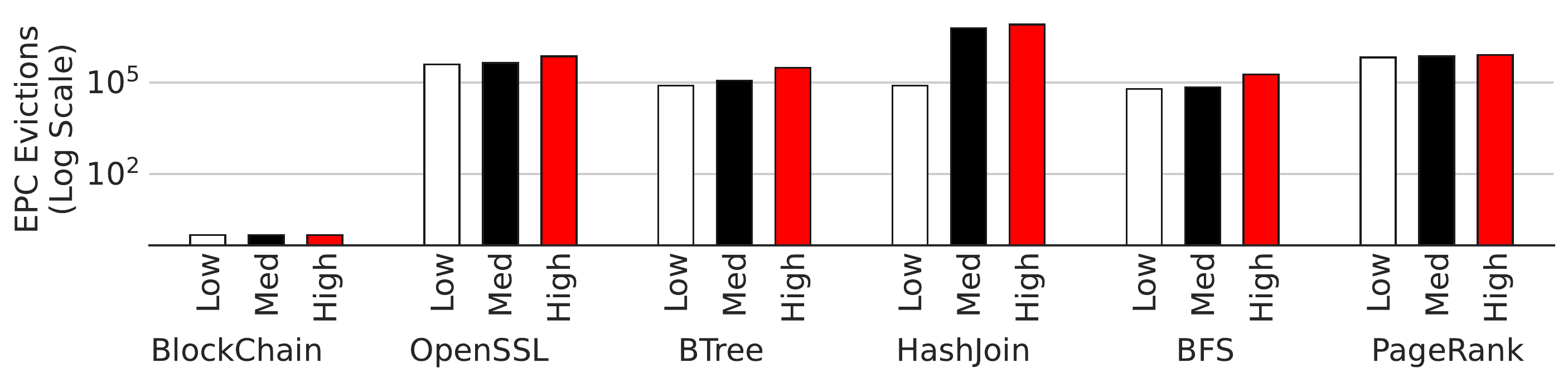}
\caption{EPC evictions in \native mode.}
\label{fig:native_overhead_native_epc_evictions}
\end{subfigure}
\caption{Performance impact of \sgx on applications in \native mode for different input sizes.}
\label{fig:performance_gauge_mid}
\end{figure}

\begin{figure*}[]
\centering

\begin{subfigure}[t]{.29\linewidth}
\includegraphics[width=\linewidth]{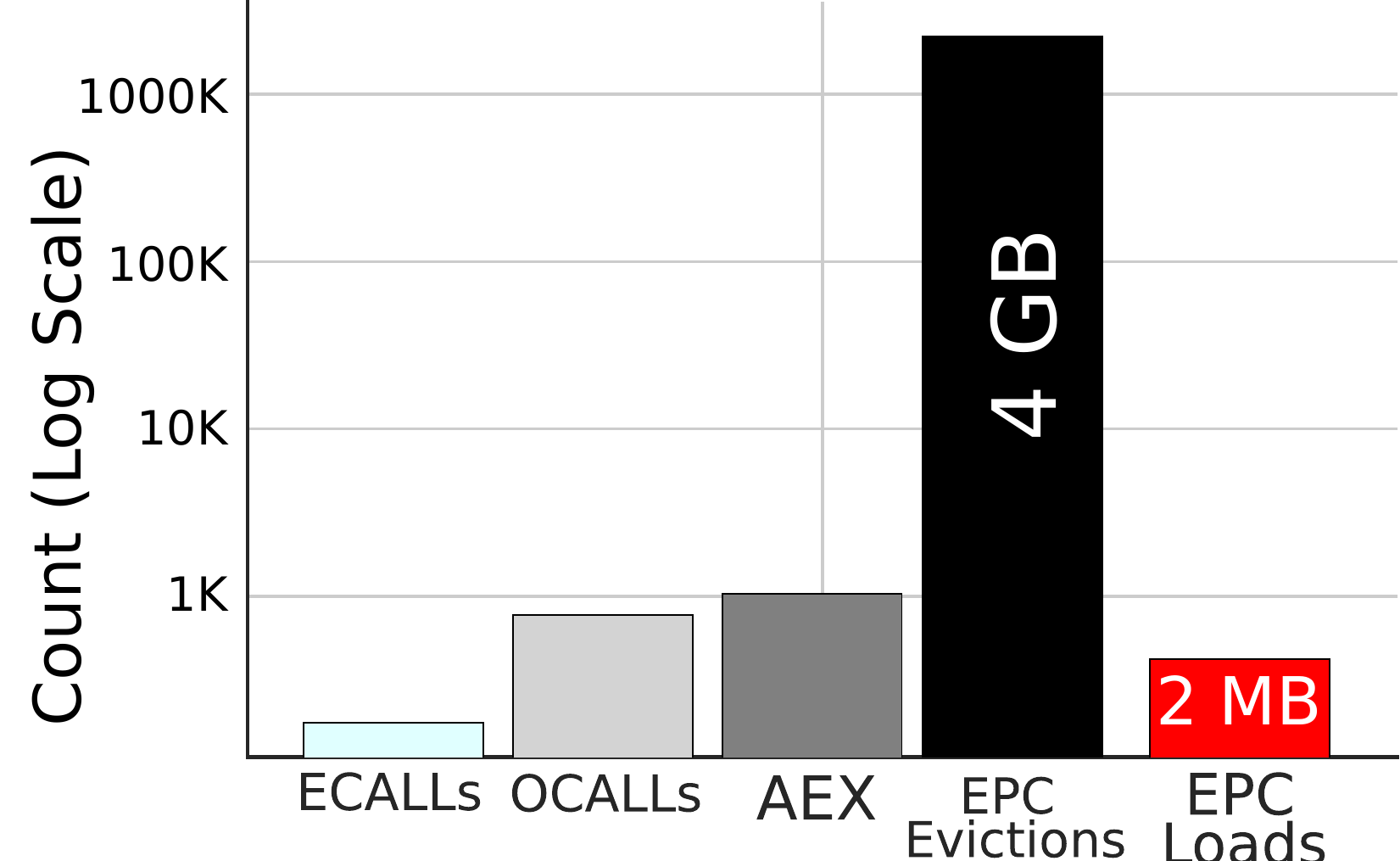}
\caption{Statistics for \graphene for an \quotes{empty} workload.}
\label{fig:graphene_empty}
\end{subfigure}
\hfil
\begin{subfigure}[t]{.69\linewidth}
\includegraphics[width=\linewidth]{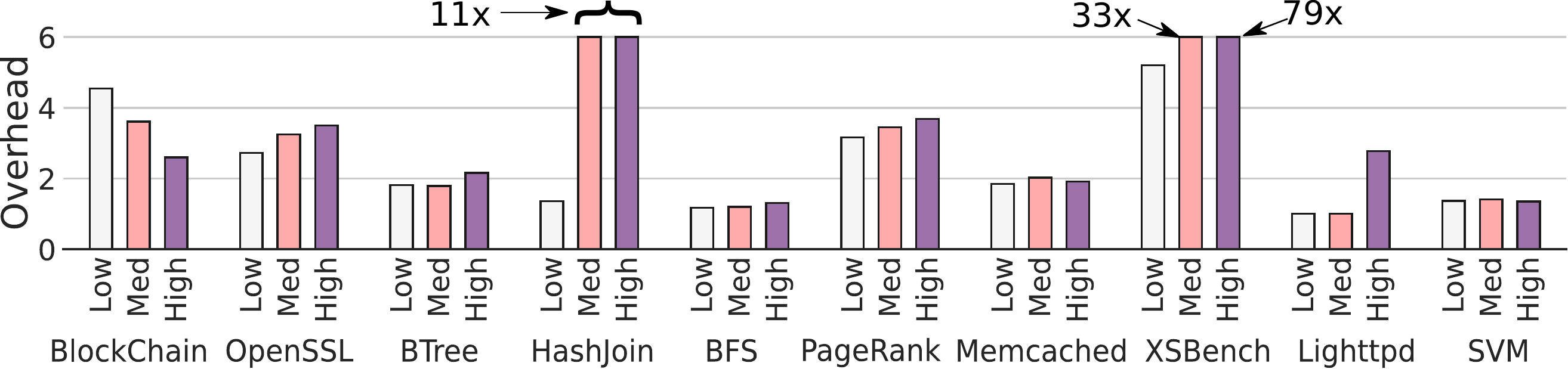}
\caption{Figure showing the overhead in the execution for the \textit{\libos} setting.}
\label{fig:graphene_overhead_sftime_mod}
\end{subfigure}

\begin{subfigure}[t]{.69\linewidth}
\includegraphics[width=\linewidth]{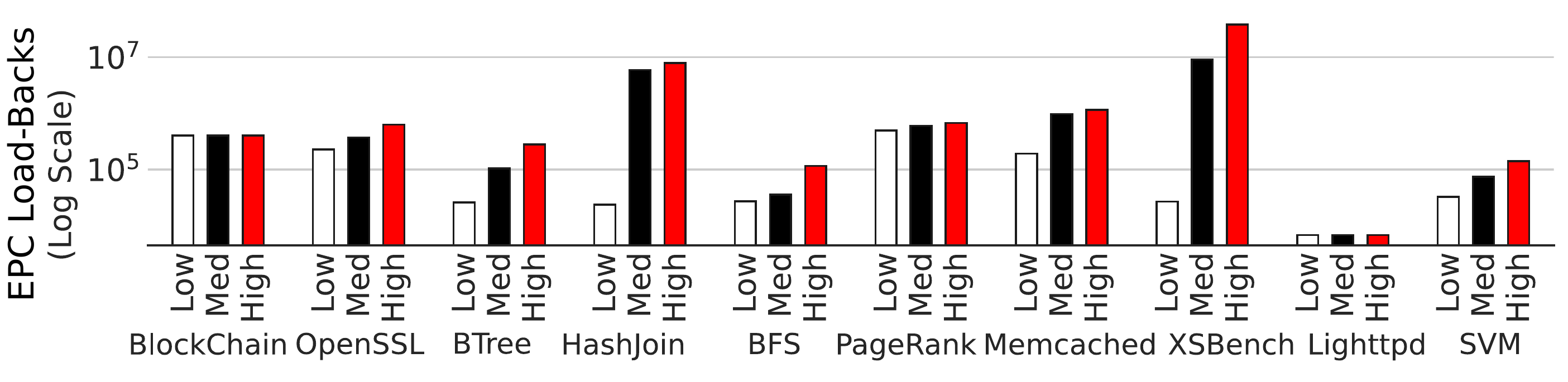}
\caption{Figure showing total EPC page reloads for the \textit{\libos} setting.}
\label{fig:graphene_overhead_graphene_ELDU_adjusted}
\end{subfigure}
\hfil
\begin{subfigure}[t]{.29\linewidth}
\includegraphics[width=\linewidth]{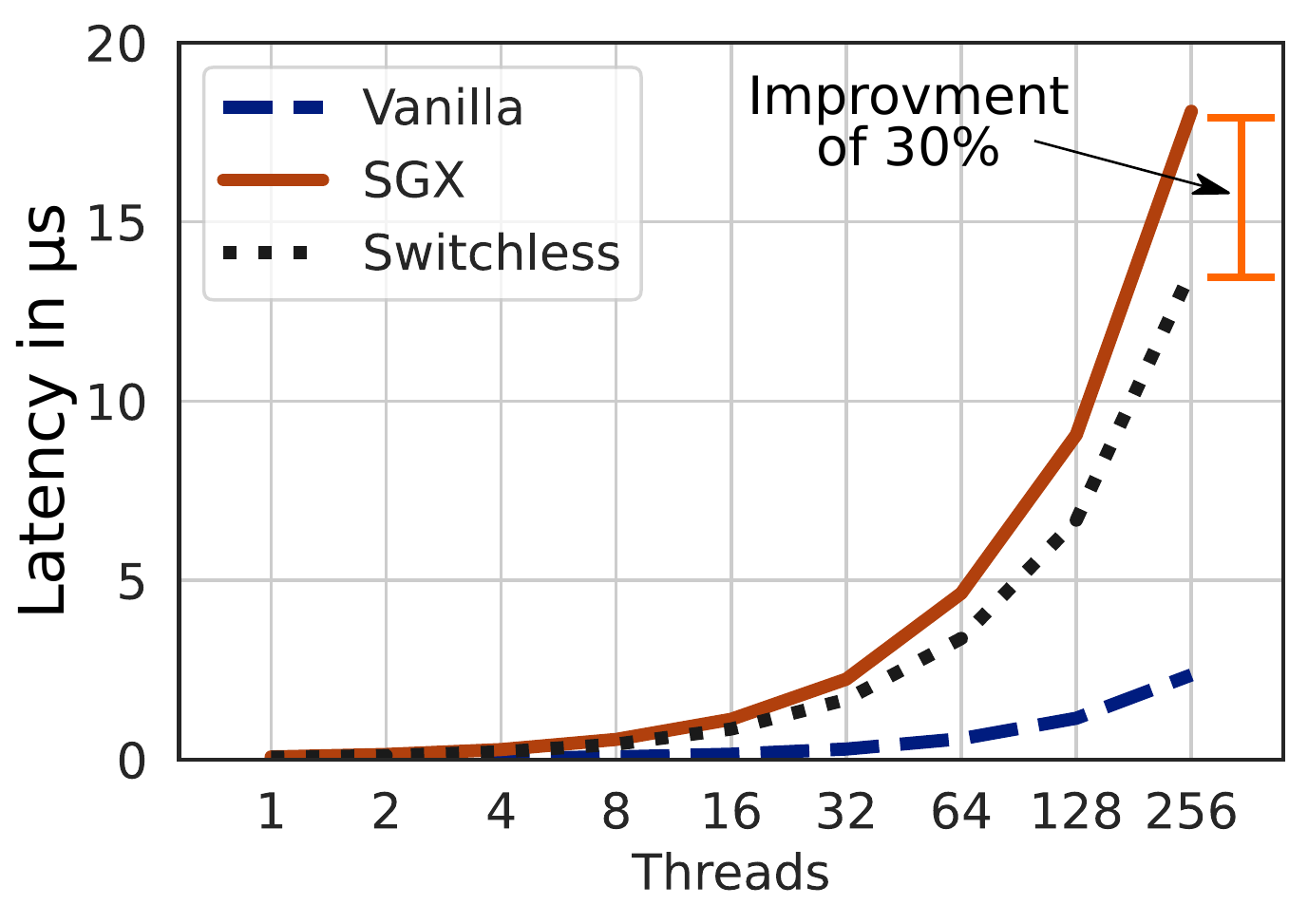}
\caption{The latency of \lighttpd improves when using the {\em switchless} mode }
\label{fig:switchless}
\end{subfigure}

\caption{Performance impact of \graphene on workloads in \methodname.}
\label{fig:performance_gauge_mid_graphene}
\end{figure*}

\subsection{\native mode Performance}
\label{sec:native_perfomance}
Here, we evaluate the performance overhead of running an application in the \native mode as compared to the \vanilla mode with different input sizes.
As shown in Figure ~\ref{fig:native_overhead_sftime_mod}, the performance overhead increases by up to $8.8\times$ as we go from the \low to the \medium setting, and by up to $1.4\times$ from the \medium to the \high setting.

As shown in Figure ~\ref{fig:native_overhead_native_epc_evictions}, the total number of EPC evictions increase by up to $75\times$ when the input size is increased from \low to \medium. On further increasing the input from \medium to \high, the total number of EPC evictions increases by up to $2.6\times$.
As already explained in Section ~\ref{sec:background}, the TLB entries of an enclave are flushed before a transition to the unsecure region due to security reasons.
Hence, as we increase the size from \low to \medium,  dTLB misses increase by up to $79\times$, and by up to $2.6\times$ as we go
from \medium to \high.
Due to this, the total walk cycles increase by up to $79\times$ (\low to \medium), and by up to $2.6\times$ while going from \medium to \high.
Consequently, the total stall cycles increase by up to $43\times$ (\low to \medium), and by up to $5.4\times$ while going from \medium to \high.

\noindent
\textbf{Summary:} As we approach the EPC size (\low to \medium), there is a sudden rise in all the paging and TLB-related performance counters. However, going beyond the EPC size (\medium to \high) does not affect the performance to that extent.
We present a detailed discussion for all the workloads in Appendix~\ref{appendix:native_workloads} and impact of the counters on the workloads' performance in Appendix~\ref{appendix:benchmark_execution_guide}.

\subsection{\libos Mode Performance}
Here, we evaluate the performance impact of \graphene. 

\subsubsection{\textbf{\graphene Overhead}}
We first characterize the overhead of just \graphene using an \quotes{empty} (\texttt{return 0;}) workload.
As shown in Figure ~\ref{fig:graphene_empty}, in this setup, \graphene performs $\approx$300 {\ecall}s, $\approx$1000 {\ocall}s, and $\approx$1000 {\aex} exits. During this time, total EPC evictions are $\approx$ 1\,M. However, out of these 1\,M evicted EPC pages, only $\approx$ 700 pages (2\,MB) are loaded back.  

The reason for the unusually high number of EPC evictions is the enclave size property, which is set to 4\,GB. 
As prior to executing an enclave, \sgx completely loads it in the EPC to calculate its signature~\cite{everything_sgx_virtual}. Doing so for a 4\,GB enclave will cause 1\,M EPC faults (1\,M*4\,KB=4\,GB). 
Lowering the value of the property \quotes{enclave-size} reduces the EPC evictions but worsens the performance by up to 4$\times$, even for the workloads with a small memory footprint such as Blockchain.
All these EPC evictions are done at the beginning of the execution, i.e., while initializing \graphene. We do not count this time in the execution time of a workload running on it (see Appendix ~\ref{appendix:graphene_startup_overhead}). 

\noindent
\textbf{Performance}
As shown in Figure~\ref{fig:graphene_overhead_sftime_mod}, the performance overhead increases by up to $8.7\times$ while going from \low to \medium, and by up to $2.7\times$ while going from \medium to \high.
As shown in Figure~\ref{fig:graphene_overhead_graphene_ELDU_adjusted}, the total number of EPC load-backs (page brought back to the EPC from the untrusted memory) increases by up to $341\times$ when the input size increases from \low to \medium, and by up to $4.1\times$ on further increasing the input size from \medium to \high.
The total number of dTLB misses increases by up to $18.6\times$ as we go from \low to \medium, and by up to $1.5\times$ as we go from \medium to \high.
Due to this, the total number of walk cycles increases by up to $11\times$ as we go from \low to \medium, and by up to $2.7\times$ when 
we go from \medium to \high.
Consequently, the total number of stall cycles increases by up to $5.11\times$ and $2.17\times$, respectively.

\noindent
\textbf{Summary:} Similar to \native mode, as we approach the EPC size (\low to \medium), there is a sudden increase in all the TLB and paging-related performance counter values. However, going beyond the EPC size (\medium to \high) does not affect the performance as much. We discuss the I/O related overheads with \graphene in Appendix~\ref{appendix:io}.

\subsection{\native Mode vs \libos Mode}
Here, we compare the performance of the \native and \libos modes (see Table ~\ref{tab:results_overview}). We observe that as we increase the input size, the performance overhead of \libos as compared to the \native mode starts decreasing. The overall overhead reduces by $12.6\%$ when we increase the input setting from \low to \high or \medium to \high.
The total number of dTLB misses comes down by $18\%$ and $25\%$, when the input is increased from \low to \medium and \medium to \high, respectively.
In the same setting, the total walk cycles comes down by 21\% and 25\%, stall cycles by 4\% and 25\%, LLC misses by 1\% and 21\%, and EPC evictions by 9\% and 33\% when we increase the input size from \low to \medium and \medium to \high, respectively.

\noindent
\textbf{Summary:} On increasing the workload size, the overhead of \graphene starts decreasing, and eventually, approaches that of the \native mode.
%

\subsection{Switchless Mode}
\label{sec:switchless_mode}
To reduce the cost of an \ocall, \intelsgx supports a switchless mode of operation where it leverages multiple cores of a modern system to make an \ocall without exiting an enclave -- thus preventing a TLB flush. In this case, a set of  threads (proxy threads) on  dedicated cores are used to handle the {\ocall}s. Here, the parameters of an \ocall and other relevant data are sent to a {\em proxy thread} running on another core using an unsecure shared memory channel. The proxy thread reads the request and performs the operation. Once the operation is finished, the results are written to the shared memory region; these results are subsequently read by the enclave that issued the request. This is a standard pattern and is used to to hide
the overheads of system calls as well in regular operating systems.

We configured \graphene to use 8 cores for handling \ocall requests from enclaves. 
In \lighttpd, this reduces the total number of dTLB misses by 60\% thus improving the latency by 30\%, as compared to the default implementation of \ocall(see Figure~\ref{fig:switchless}).

\section{Conclusion}
\label{sec:conclusion}
We introduced SGXGauge, a benchmark suite for Intel
SGX that captures a holistic view of the performance of
applications running in such TEEs – this includes the impact
of the EPC memory. SGXGauge contains diverse benchmarks
that affect different components of SGX. We also performed
an evaluation of the performance of SGX in LibOS mode and
showed that there is a marked difference in behavior as the
memory footprint crosses the EPC size limit.


\newpage

\bibliographystyle{ACM-Reference-Format}
\bibliography{refs}

\appendix

\section{\intelsgx Latencies}
\label{appendix:latency}
Instrumenting an application running in the secure mode within \sgx is a non-trivial task due to the constraints imposed by \intelsgx.
\texttt{RDTSC} instructions~\cite{rdtsc,rdtsc_micro}, which are generally used to measure the cycles taken by an operation, are not allowed in \sgx. Fortunately, the \intelsgx driver code does not execute in \intelsgx and thus can be easily instrumented.

We measured the latencies of the core \intelsgx operations: allocating a page (\texttt{sgx\_alloc\_page()}), evicting a page (\texttt{sgx\_ewb()}), loading back a page (\texttt{sgx\_eldu()}), and handling a page fault in \sgx(\texttt{sgx\_do\_fault()}). We use the \textit{ftrace} tool~\cite{ftrace} for this purpose.
\sgx uses the \texttt{EWB} instruction to evict a page from the EPC and the \texttt{ELDU} instruction to load it back.
While evicting  an EPC page, first its MAC is calculated and then the page is encrypted. While loading back, the page is decrypted and its integrity is checked using the MAC~\cite{vault}.

\begin{figure}[!ht]
\centering
\includegraphics[width=.7\linewidth]{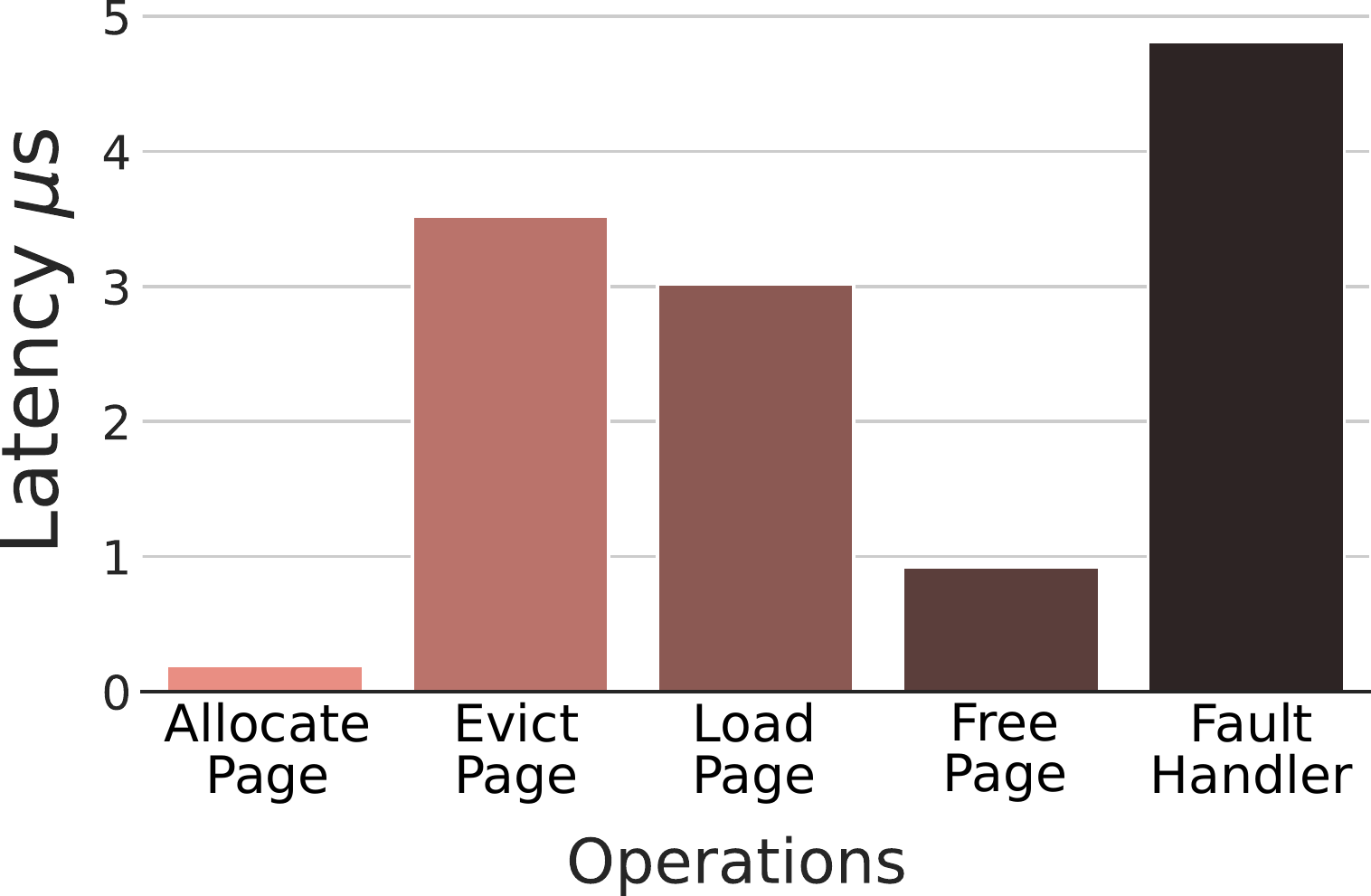}
\caption{Latency of core operations in \intelsgx.}
\label{fig:latency}
\end{figure}

These functions are highly optimized with latencies in the range of a few micro-seconds. Figure~\ref{fig:latency} reports the  mean of 40K+ samples. The latency of evicting an EPC page is $16\%$ more than loading back an EPC page. \sgx evicts pages in a batch that is typically 16 pages. However, during a fault, a single page is loaded back.

\section{\native Mode Performance}
\label{appendix:native_workloads}
\begin{figure}
\centering
\begin{subfigure}[t]{.565\linewidth}
\includegraphics[width=\linewidth]{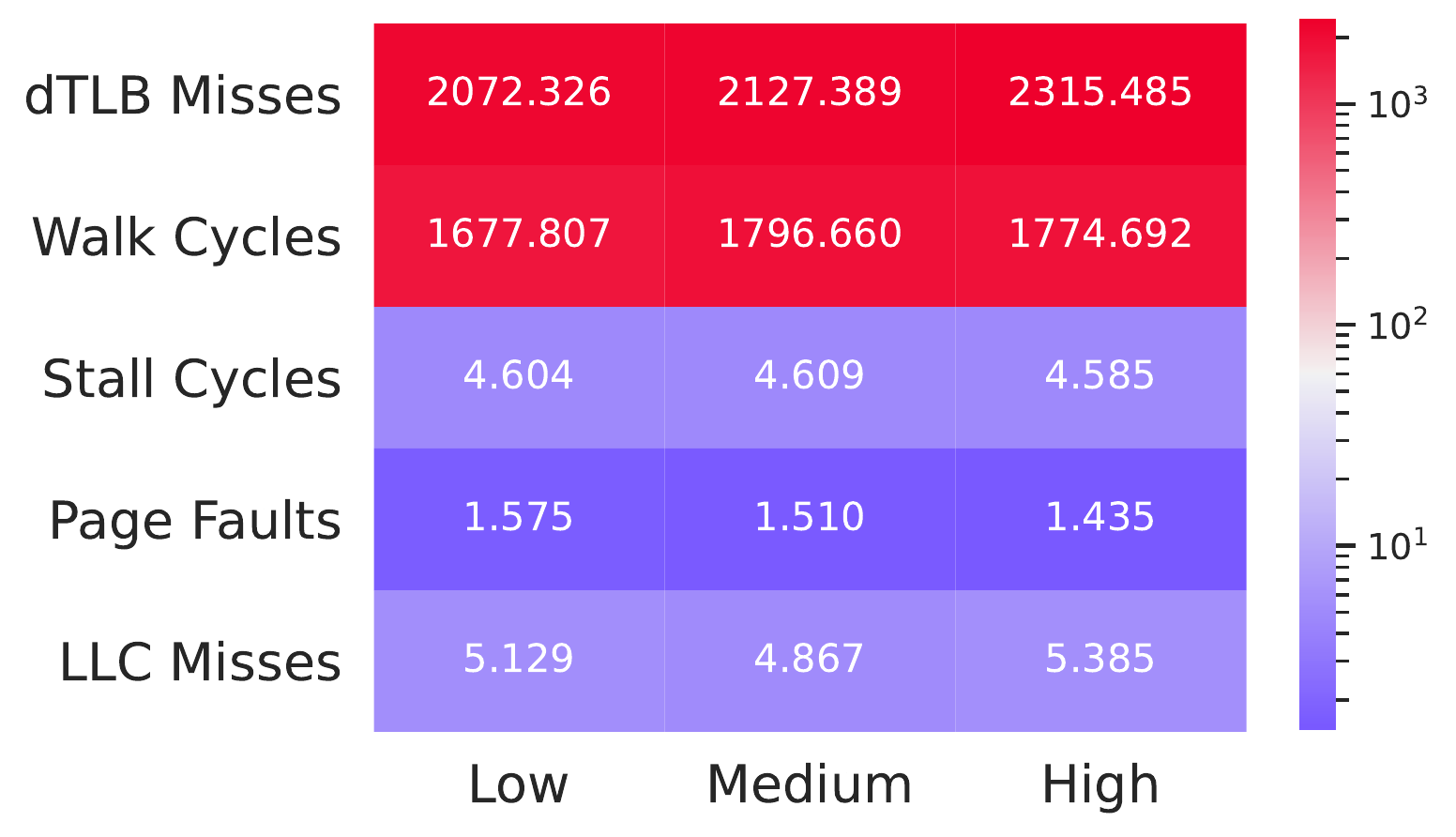}
\caption{\blockchain}
\label{fig:sn_blockchain}
\end{subfigure}
\hfil
\begin{subfigure}[t]{.425\linewidth}
\includegraphics[width=\linewidth]{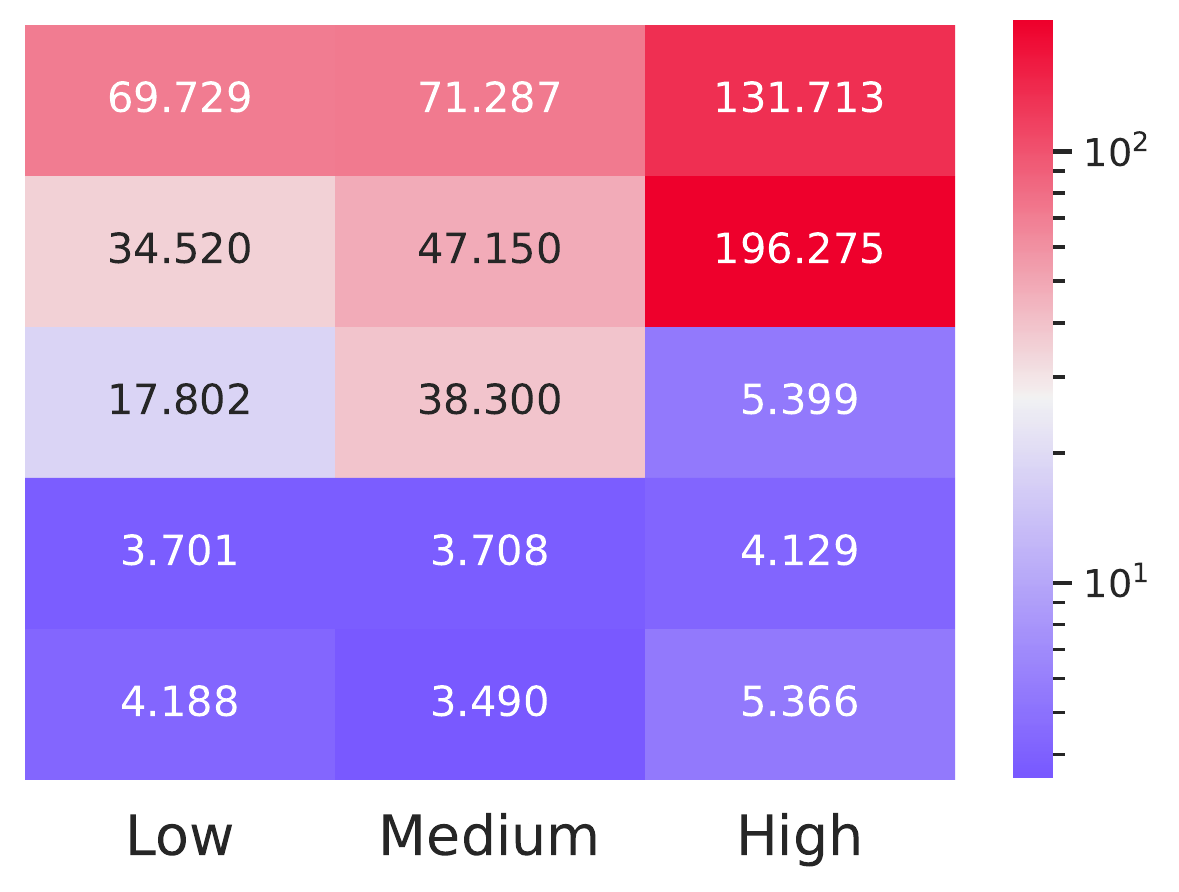}
\caption{\openssl}
\label{fig:sn_openssl}
\end{subfigure}

\begin{subfigure}[t]{.565\linewidth}
\includegraphics[width=\linewidth]{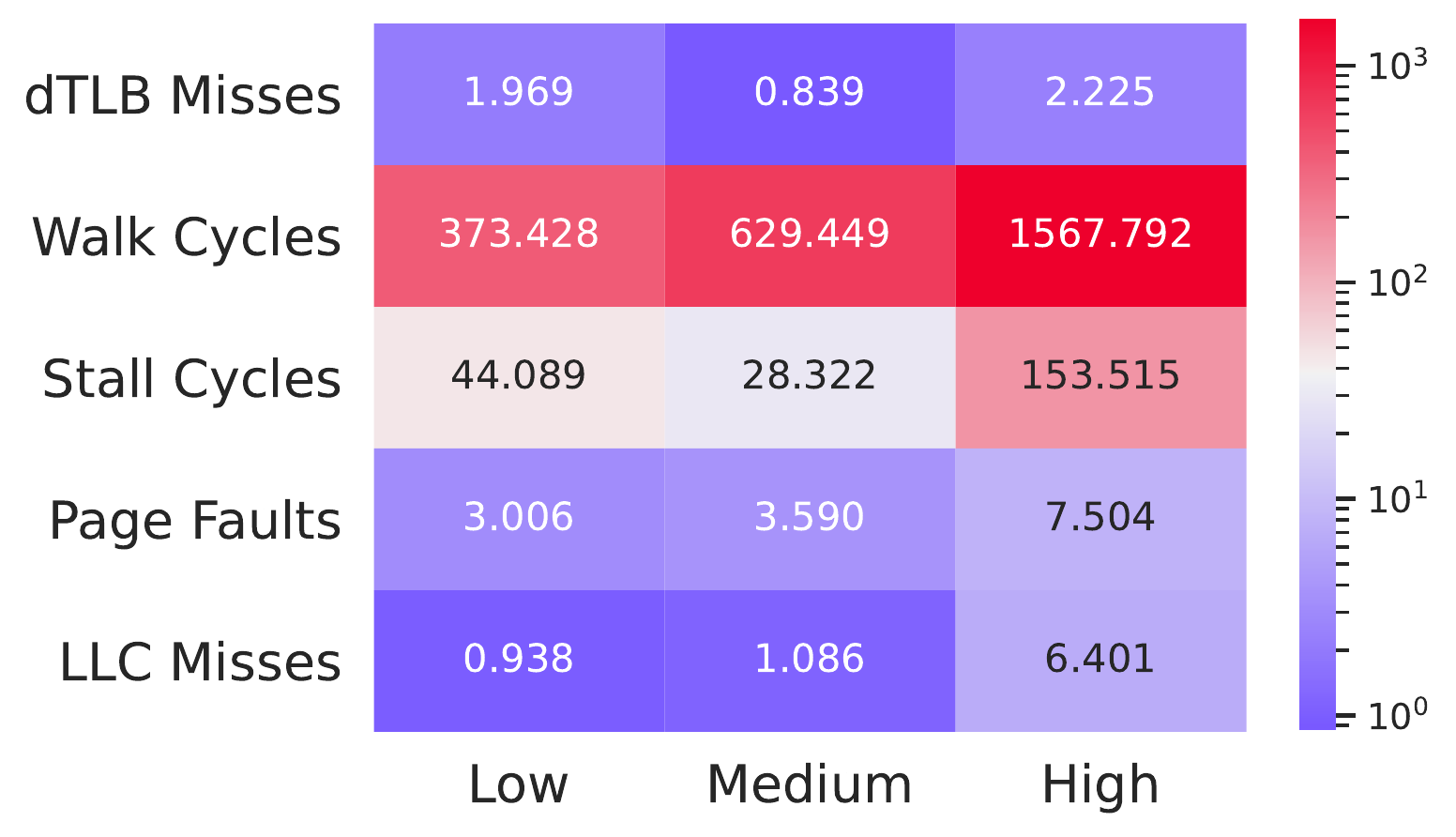}
\caption{\btree}
\label{fig:sn_btree}
\end{subfigure}
\hfil
\begin{subfigure}[t]{.425\linewidth}
\includegraphics[width=\linewidth]{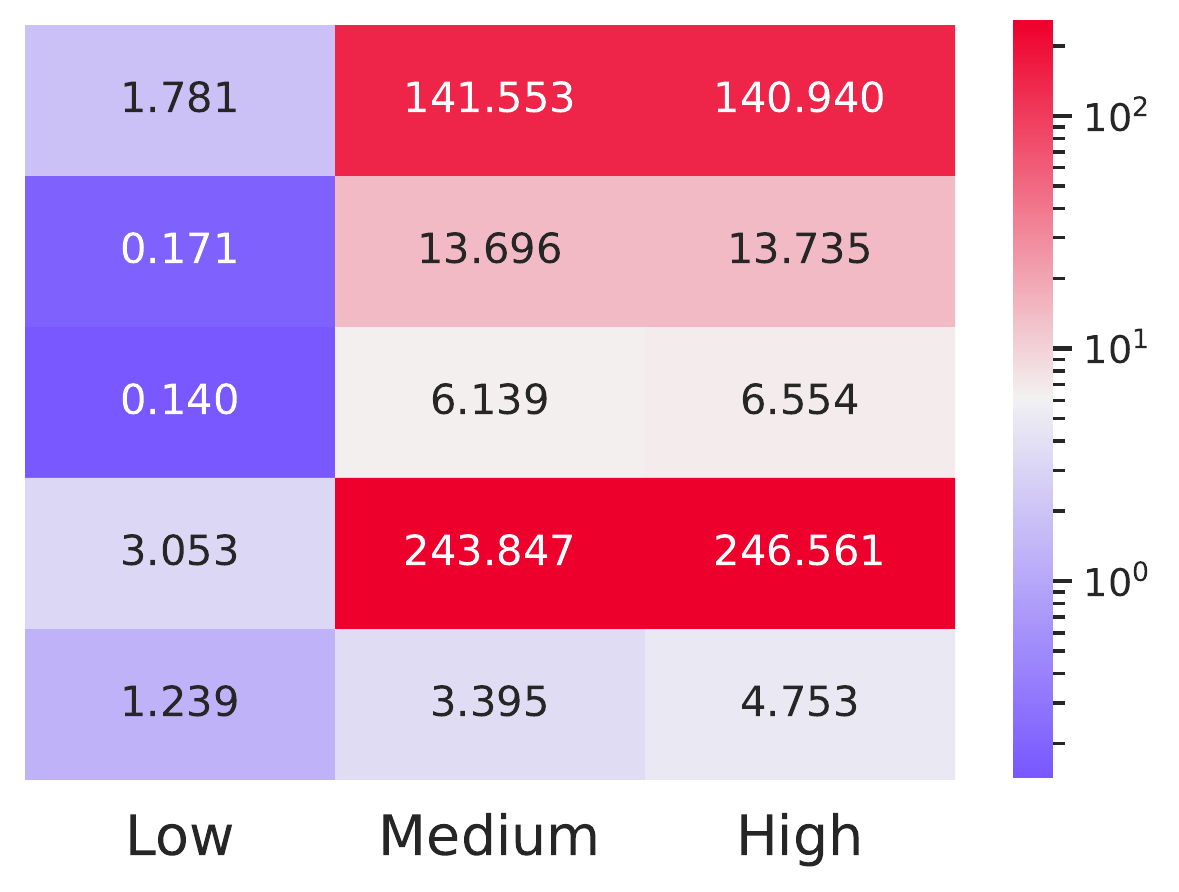}
\caption{\hashjoin}
\label{fig:sn_hashjoin}
\end{subfigure}

\begin{subfigure}[t]{.565\linewidth}
\includegraphics[width=\linewidth]{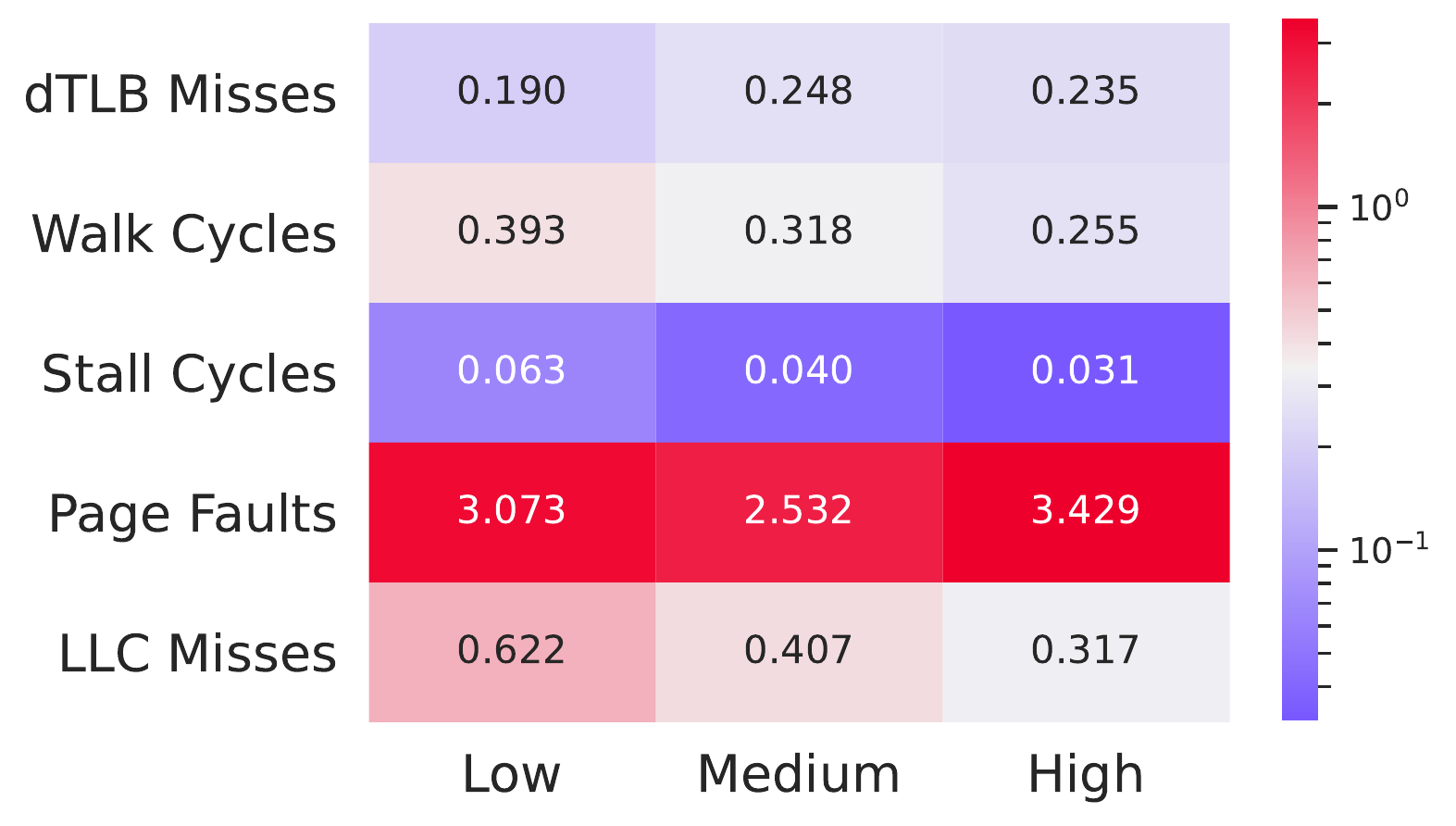}
\caption{\bfs}
\label{fig:sn_bfs}
\end{subfigure}
\hfil
\begin{subfigure}[t]{.425\linewidth}
\includegraphics[width=\linewidth]{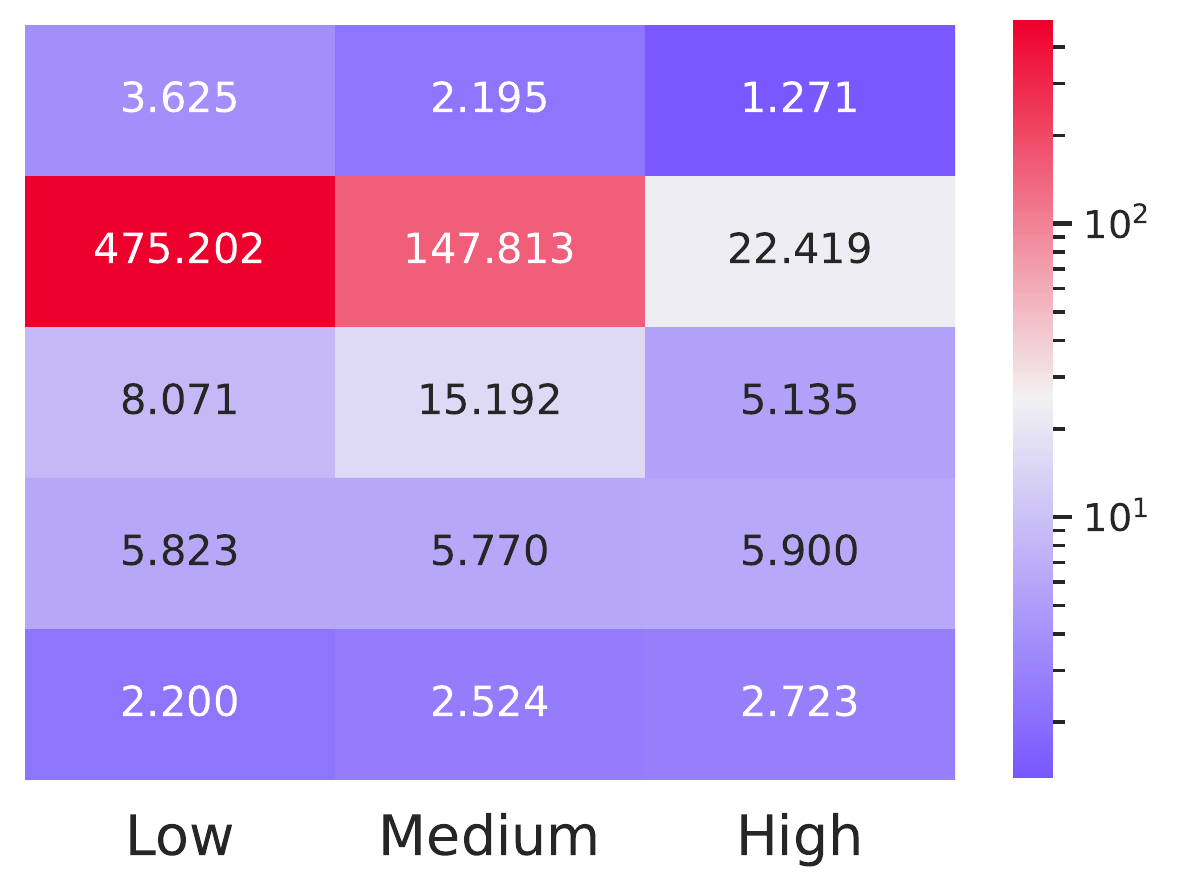}
\caption{\pagerank}
\label{fig:sn_pagerank}
\end{subfigure}

\caption{Overheads for the workloads when executing in the \native mode w.r.t the \vanilla mode.}
\label{fig:sn_heatmap}
\end{figure}

Here, we discuss the overheads in the 6 of workloads from \methodname while executing in the \native mode. We use the change in the most relevant hardware performance counters for this discussion. A heat-map of these counters is shown in Figure~\ref{fig:sn_heatmap}.

\subsection{\blockchain}
As seen in Figure~\ref{fig:sn_blockchain}, while executing in the \native mode, the total number of dTLB misses is ~$\approx$ 2000$\times$ more than the \vanilla mode. This is because in our implementation of \blockchain the hash function is protected inside an enclave. This is called via {\ecall}s from many threads that are executing in the unsecure region. This results in many enclave transitions and thus many TLB flushes. These TLB entries are to be populated for every \ecall by a page table walk. Hence, we see a similar increase in the number of walk cycles.

With 16 threads in the \low setting there are $\approx$ 3,133\,K {\ecall}s, for \medium $\approx$ 4,831\,K {\ecall}s, and for the \high setting there are $\approx$ 8,944\,K {\ecall}s.

\subsection{\openssl}
In \openssl, the number of EPC evictions increases from 389\,K to 433\,K to 721\,K, as we increase the input size from  \low to \medium and \high, respectively. Due to this, in the \high setting, the total number of enclave exits increases, thus increasing the total number of dTLB misses by 131$\times$ and walk cycles by 196$\times$ w.r.t. the \vanilla mode. 

\subsection{\btree}
In \btree, the total number of EPC evictions increases from 79\,K to 116\,K to 305\,K, when we move from the \low to \medium and then to the \high input setting, respectively. However, here the total number of dTLB misses increase by only 2.2$\times$ only in the \high setting. 
This is because the total number of dTLB misses is dominated by the number of page faults caused by the workload. To serve a page fault, an enclave performs an asynchronous exit (AEX), which also causes a TLB flush.
The total number of page faults increases from 3$\times$ in the \low setting to 7.5$\times$ in the \high setting, and the total LLC misses increase from $\approx$1 in the \low setting to $\approx$6.4 in the \high setting.

\subsection{\hashjoin}
In \hashjoin, on increasing the input size we observe an increase in almost all of the performance counters. Most notably, the total number of page faults and dTLB misses increase by $\approx$246$\times$ and $\approx$140$\times$ over \vanilla mode in the \high input setting, respectively. 
This is due to the characteristics of the workload. A typical hashjoin operation incurs many cache misses and stall cycles~\cite{hashjoin_perf}.

\subsection{\bfs}
In \bfs, the total number of page faults increase by 3$\times$ as compared to the \vanilla mode. However, we do not observe a large impact with the increase in the input size. This is because of the inherent locality in the workload.

\subsection{\pagerank}
In \pagerank, we observe a decrease in the total number of walk cycles on increasing the input size. This is because dTLB misses also go down with an increasing input size. The main reason for this is the nature of the workload. In the \vanilla mode (not shown in the Figure), the number of dTLB misses increases by 3.6$\times$ when we increase the input size from the \low to the \high setting. Hence, the nature of the workload dominates the total number of misses, hiding the extra misses caused due to \sgx.

\section{Counter Impact on Performance}
\label{appendix:benchmark_execution_guide}

\intelsgx provides a way to execute an application securely on a remote machine, although with some limitations. Researchers are working on developing methods to circumvent these limitations. 
Different solutions might affect the components of \intelsgx differently. Here we provide a generic approach for the developer to select correct benchmarks from \methodname as per the requirement.

As pointed out in prior work and observed by us in our experiments, when a benchmark reserves more memory than the EPC size it suffers a slowdown. Usage of more memory than the EPC may impact the total number of dTLB misses, LLC misses, walk cycles, stall cycles, and EPC-Evictions.
We rank the metrics in the order of importance for each of the workloads in \methodname. We use linear regression~\cite{multiple_linear_regression} for this purpose. Linear regression predicts the execution time given these metrics as input. While doing so, it assigns coefficients to these metrics. The magnitude of these coefficients is correlated with the importance of that metric in determining the execution time (see Table~\ref{tab:linear_regression}).

\begin{table}[!ht]
\centering
\footnotesize
\caption{Table showing the most important hardware performance counter that determines the performance of each workload (shown in {\bf bold}). LLC refers to the last level cache.}
\label{tab:linear_regression}
{%
\begin{tabular}{|l|p{.8cm}|p{.7cm}|p{.8cm}|p{.8cm}|p{.8cm}|p{.9cm}|}
\hline
Workloads      & Walk cycles & Stall cycles & Page faults & dTLB misses & LLC misses & EPC evictions \\ \hline
\rowcolor{LightCyan} \multicolumn{7}{|c|}{\native mode} \\ \hline
{Blockchain} & \textbf{0.33}         & 0.32          & 0.01         & 0           & 0.32        & 0             \\ \hline
{OpenSSL}    & 0.08         & 0.12          & 0.17         & 0           & \textbf{0.21 }       & 0.14           \\ \hline
{BTree}      & \textbf{0.27}     & $-$0.11     & 0.22     & $-$0.05    & 0.22    & 0.11       \\ \hline
{HashJoin}   & \textbf{0.21}     & 0.16      &\textbf{ 0.21}     & $-$0.03    & \textbf{0.21 }   & \textbf{0.21  }     \\ \hline
{BFS}        & 0.10     & 0.17      & 0.18     & 0.21     & 0.09    &\textbf{ 0.22 }      \\ \hline
{PageRank}   & 0.44     & \textbf{$-$0.54}     & 0.05     & 0.33     & 0.65    & 0.04       \\ \hline
\rowcolor{LightCyan} \multicolumn{7}{|c|}{\libos mode} \\ \hline
{Memcached}  & 0.03        & 0.04         & 0.09        & \textbf{0.15}        & 0.13       & 0.09          \\ \hline
{XSBench}    & 0.16        & 0.17         & 0.17        &\textbf{ 0.18}        & 0.16       & 0.17          \\ \hline
{Lighttpd}   & 0.18        & 0.19         & 0           & 0.09        &\textbf{ 0.26}       & 0             \\ \hline
{SVM}        & 0.09        & \textbf{0.60 }        & 0.27        & 0.09        & 0.31       & $-$0.03         \\ \hline
\end{tabular}%
}
\end{table}

We can conclude that most of the time paging and TLB-related counters are the most correlated with the performance. LLC misses are mostly an important factor in \openssl.

\section{\graphene start-up overhead}
\label{appendix:graphene_startup_overhead}

Here, we discuss the overhead in initializing \graphene.
\begin{figure}[!ht]
\centering
\includegraphics[width=.9\linewidth]{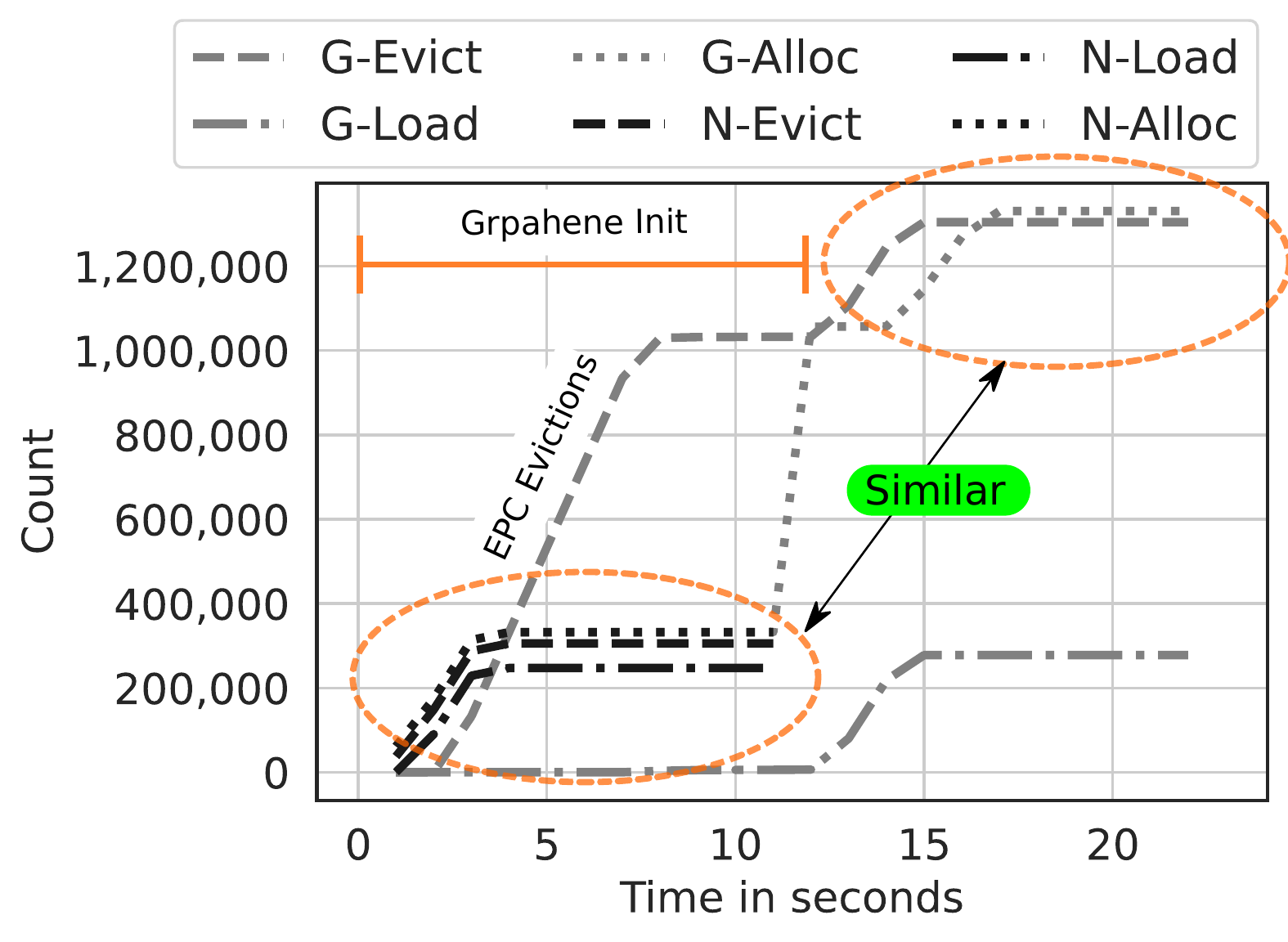}
\caption{Figure showing the performance counter values for EPC page allocation, eviction, and load-back during the execution phase of \btree in the \native mode (N-) and \libos mode (G-).}
\label{fig:graphene_alloc}
\end{figure}

\intelsgx verifies the signature of an enclave prior to its execution. To do so, it loads the entire enclave into the EPC. In \sgx v1, a heap size greater than the EPC size was not allowed, as that will not allow \sgx to load the complete enclave in the EPC. Since \sgx v2, a heap size greater than the EPC is allowed. \sgx transparently evicts and loads pages as per its requirement.

Figure~\ref{fig:graphene_alloc} shows the allocation, eviction, and load back of EPC pages in \native and \libos modes for a representative workload, \btree. This pattern remains the same across other workloads also. \sgx first calculates the signature of the enclave that causes the initial EPC evictions. Note that EPC pages are allocated after the verification is done. After that, the EPC pattern of \graphene is the same as that of the \native mode. 

\intelsgx recommends setting the enclave size as per the maximum requirement of the application. However, in our experiments we observed additional overheads on setting a lower value of the enclave size in \libos mode. This is related to how \graphene initializes the enclave. We thus used an enclave size of 4\,GB for all our experiments. We do not count the \graphene start up time in the workload executing time while calculating the overheads in Section~\ref{sec:evaluation} mainly because this 
is an one-time activity and a workload can run for a very long time after its enclave is initialized. Also note in 
Figure~\ref{fig:graphene_alloc} that after the initialization phase the gray (\graphene) and black (\native) lines
converge (same behavior).

\section{What About  I/O?}
\label{appendix:io}

As mentioned before, \sgx does not support system calls, notably file system calls. An enclave needs to rely on an \ocall to read or write a file to the file system. 
In this case, by default, the data is transferred in plaintext and it is the responsibility of the developer to protect it via encryption. \sgx has a sealing feature, where the data can be encrypted using the \textit{sealing} enclave~\cite{intelsgxexplained}. The sealing enclave is an Intel-authored enclave that is part of the Intel SDK. It can \quotes{seal} or encrypt data using a platform dependent hardware key. The sealed data can only be \quotes{unsealed} or decrypted on the same platform, and optionally, it can be configured to be decrypted only by the same enclave that encrypted it. 

\begin{figure}[!ht]
\centering
\begin{subfigure}[!b]{.49\linewidth}
\includegraphics[width=\linewidth]{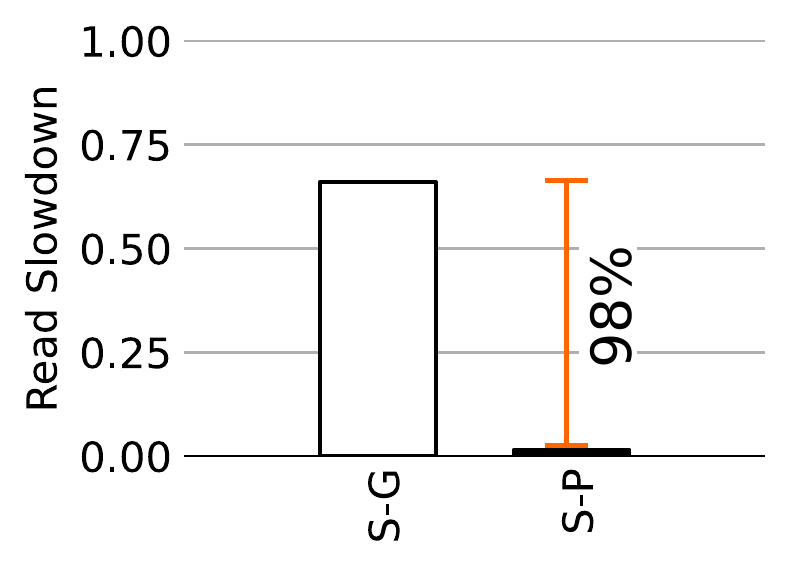}
\caption{Read performance}
\end{subfigure}
\begin{subfigure}[!b]{.49\linewidth}
\includegraphics[width=\linewidth]{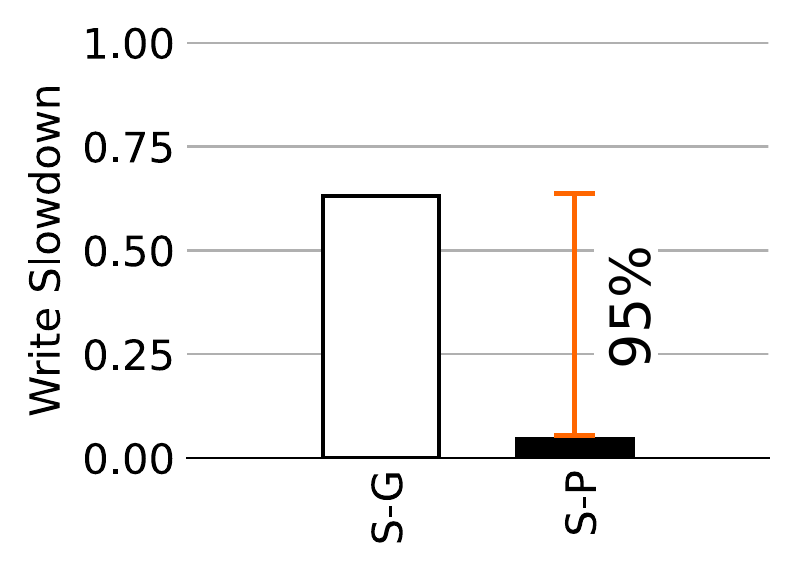}
\caption{Write performance}
\end{subfigure}

\begin{subfigure}[!b]{.49\linewidth}
\includegraphics[width=\linewidth]{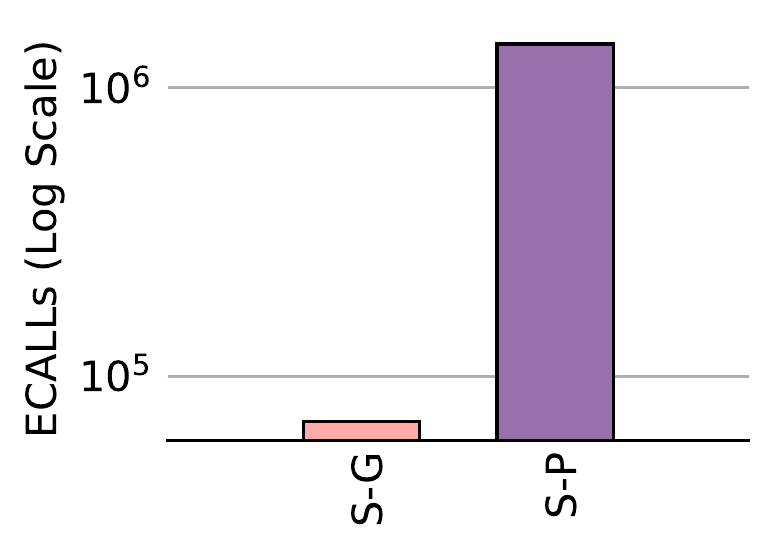}
\caption{\ecall overhead}
\label{fig:iozone_ecall}
\end{subfigure}
\begin{subfigure}[!b]{.49\linewidth}
\includegraphics[width=\linewidth]{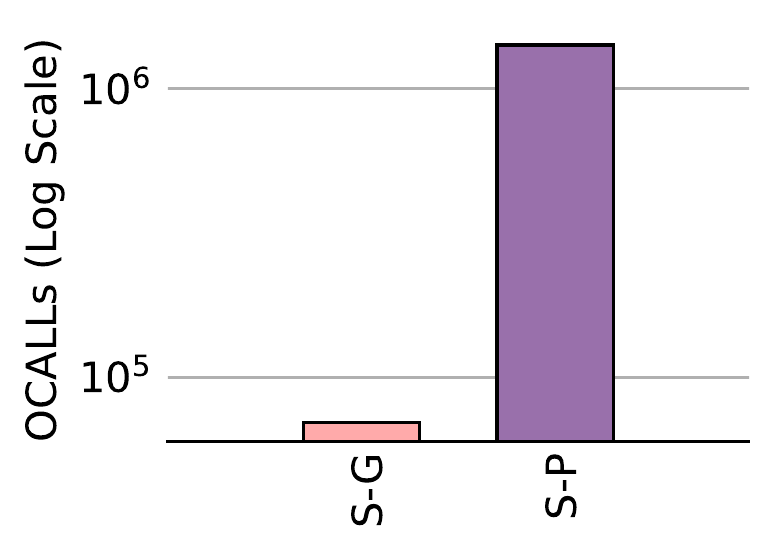}
\caption{\ocall overhead}
\label{fig:iozone_ocall}
\end{subfigure}

\caption{The I/O overhead with \graphene (S-G) and \graphene with protected files (S-P). \iozone: reading and writing 1\,GB of data with 4\,M blocks.}
\label{fig:iozone_performance}
\end{figure}

Library operating systems support file system operations by transparently capturing these calls and handling them either via an \ocall or via a parallel, proxy thread executing on a different core. However, as this is not covered by \intelsgx constructs, a naive implementation will still write the data in plain text to the  file system, essentially leaking data. \graphene supports transparently encrypting files before they are written to the file system. This feature is known as the protected file system or PF~\cite{graphenepf} mode. However, as shown in Figure~\ref{fig:iozone_performance} this feature is not optimized, and the performance of an I/O intensive application can suffer by up to $98\%$ when PF is used.

We use the popular file system benchmark Iozone~\cite{iozone} to evaluate the performance of the \graphene PF system. We compare this against the \vanilla mode, and \libos mode without using the protected file setting.
\libos incurs an overhead of 33\%  and 36\% compared to the \vanilla mode for read and write operations, respectively. The overhead increases to $98\%$ and $95\%$ for read and write operations, respectively, when the protected files mode is enabled. The main reason for this is the increase in the number of {\ecall}s (see Figure~\ref{fig:iozone_ecall}) and {\ocall}s (see Figure~\ref{fig:iozone_ocall}).

The PF mode needs to be optimized to make it practical for production-quality systems.

\end{document}